\documentclass[preprint]{emulateapj}
\usepackage{apjfonts}
\usepackage{epsfig}
\usepackage{natbib}
\newcommand{\kms}{km\ s$^{-1}$}

\usepackage{natbib}
\def\apjs{{ApJS}}

\def\micron{{{$\mu m $}}}
\def\s4g{{S$^4$G}}

\def\ser{S{\'e}rsic}

\shorttitle{ETGs with Tidal Debris and their Scaling Relations in {\s4g}}
\shortauthors{Kim et al.}

\begin{document}
\title{Early Type Galaxies with Tidal Debris and their Scaling Relations in the Spitzer Survey of Stellar Structure in Galaxies ({S$^4$G})}

\author{Taehyun Kim\altaffilmark{1,2,3},
Kartik Sheth\altaffilmark{1,4,5},
Joannah L. Hinz\altaffilmark{6},
Myung Gyoon Lee\altaffilmark{3},
Dennis Zaritsky\altaffilmark{6},
Dimitri A. Gadotti\altaffilmark{2},
Johan H. Knapen\altaffilmark{7,8},
Eva Schinnerer\altaffilmark{9},
Luis C. Ho\altaffilmark{10},
Eija Laurikainen\altaffilmark{11,12},
Heikki Salo\altaffilmark{11},
E. Athanassoula\altaffilmark{13},
Albert Bosma\altaffilmark{13},
Bonita de Swardt\altaffilmark{14},
Juan-Carlos Mu\~noz-Mateos\altaffilmark{1},
Barry F. Madore\altaffilmark{10},
S\'ebastien Comer\'on\altaffilmark{15},
Michael W. Regan\altaffilmark{16},
Kar\'in Men\'endez-Delmestre\altaffilmark{17},
Armando Gil de Paz\altaffilmark{18},
Mark Seibert\altaffilmark{10},
Jarkko Laine\altaffilmark{11},
Santiago Erroz Ferrer\altaffilmark{7,8},
Trisha Mizusawa\altaffilmark{1,4,5}
}

\altaffiltext{1}{National Radio Astronomy Observatory / NAASC, 520 Edgemont
 Road, Charlottesville, VA 22903}
\altaffiltext{2}{European Southern Observatory, Casilla 19001, Santiago 19, Chile; tkim@eso.org}
\altaffiltext{3}{Astronomy Program, Department of Physics and Astronomy, Seoul National University, Seoul 151-742, Korea}
\altaffiltext{4}{Spitzer Science Center, 1200 East California Boulevard, Pasadena, CA 91125}
\altaffiltext{5}{California Institute of Technology, 1200 East California Boulevard, Pasadena, CA 91125}
\altaffiltext{6}{University of Arizona, 933 N. Cherry Ave, Tucson, AZ 85721}
\altaffiltext{7}{Instituto de Astrof\'\i sica de Canarias, E-38200 La Laguna, Tenerife, Spain}
\altaffiltext{8}{Departamento de Astrof\'\i sica, Universidad de La Laguna, E-38205 La Laguna, Tenerife, Spain}
\altaffiltext{9}{Max-Planck-Institut f{\"u}r Astronomie, K{\"o}nigstuhl 17, 69117 Heidelberg, Germany}
\altaffiltext{10}{The Observatories of the Carnegie Institution for Science, 813 Santa Barbara Street, Pasadena, CA 91101, USA }
\altaffiltext{11}{Division of Astronomy, Department of Physical Sciences, University of Oulu, Oulu, FIN-90014, Finland}
\altaffiltext{12}{Finnish Centre of Astronomy with ESO (FINCA), University of Turku, V{\"a}is{\"a}l{\"a}ntie 20, FI-21500, Piikki{\"o}, Finland}
\altaffiltext{13}{Laboratoire d'Astrophysique de Marseille (LAM), UMR6110, Universit{\'e} de Provence/CNRS, Technop{\^{o}}le de
Marseille Etoile, 38 rue Fr{\'e}d{\'e}ric Joliot Curie, 13388 Marseille C{\'e}dex 20, France}
\altaffiltext{14}{South African Astronomical Observatory, Observatory, 7935 Cape Town, South Africa}
\altaffiltext{15}{Korea Astronomy and Space Science Institute, 61-1 Hwaamdong, Yuseong-gu, Daejeon 305-348, Republic of Korea}
\altaffiltext{16}{Space Telescope Science Institute, 3700 San Martin Drive, Baltimore, MD 21218}
\altaffiltext{17}{Universidade Federal do Rio de Janeiro, Observat{\'o}rio do Valongo, Ladeira Pedro Ant{\^{o}}nio, 43, CEP 20080-090, Rio de Janeiro, Brazil}
\altaffiltext{18}{Departamento de Astrof\'\i sica, Universidad Complutense de Madrid, Madrid 28040, Spain}

\begin{abstract}
Tidal debris around galaxies can yield important clues on their evolution. We have identified tidal debris in 11 early type galaxies (T $\le $ $0$) from a sample of 65 early types drawn from the {\em{Spitzer}} Survey of Stellar Structure in Galaxies (\s4g). The tidal debris includes features such as shells, ripples and tidal tails. A variety of techniques, including two-dimensional decomposition of galactic structures, was used to quantify the residual tidal features. The tidal debris contributes $\sim$3 -- 10\%  to the total 3.6 {\micron} luminosity of the host galaxy. Structural parameters of the galaxies were estimated using two-dimensional profile fitting. We investigate the locations of galaxies with tidal debris in the Fundamental Plane and Kormendy relation. We find that galaxies with tidal debris lie within the scatter of early type galaxies without tidal features. 
Assuming that the tidal debris is indicative of recent gravitational interaction or merger, this suggests that these galaxies have either undergone minor merging events so that the overall structural properties of the galaxies are not significantly altered, or they have undergone a major merging events but already have experienced sufficient relaxation and phase-mixing so that their structural properties become similar to those of the non-interacting early type galaxies. 
\end{abstract}


\section{Introduction}
Early type galaxies (ETGs) used to be regarded as relatively simple, quiescent stellar systems undergoing passive evolution with little or no star formation. Various studies, however, indicate that the histories of ETGs may be more complex. Following the suggestion that most field elliptical galaxies may be the result of merging of disk galaxies \citep{toomre77}, numerous observational studies have shown that many ETGs have faint features or disturbances around them that are suggestive of recent interactions. In particular, many ETGs show tidal features such as ``shells'' (\citealt{malin80}), ``ripples'' (\citealt{schweizer80}) or distinctive tidal tails (\citealt{arp66}; \citealt{malin83}; \citealt{fort86}; \citealt{schweizer92}; \citealt{michard04}; \citealt{vandokkum05}; \citealt{canalizo07}; \citealt{bennert08}; \citealt{tal09}; \citealt{janowiecki10}). Although some shells are slightly bluer or redder than the underlying galaxies, most shell colors are found to be similar to those of their underlying galaxies (\citealt{fort86}; \citealt{forbes95}; \citealt{turnbull99}; \citealt{sikkema07}). On the other hand, tidally disturbed ETGs, are on average bluer than non-interacting ETGs and the more tidally disturbed, the bluer (\citealt{schweizer92}; \citealt{michard04}; \citealt{tal09}). 

There have been various studies on the formation of shells. Numerical simulations of galaxy mergers of unequal mass, mass ratio of 1/10 -- 1/100 (i.e minor mergers) produce shells (\citealt{quinn84}, \citealt{dupraz86}, \citealt{hernquist87b}, \citealt{hernquist87a}). In these simulations, shells are density waves from infalling stars of a low mass and low velocity dispersion companion galaxy during a merging event. Major merger models (mergers of comparable mass galaxies) can also form shells (\citealt{hernquist92}, \citealt{hibbard95}).  In these cases, the outer disk material of the pre-merger spiral falls into the merged remnant late, long after the inner region of the remnant has relaxed in the merging event (\citealt{hernquist92}).  
 A weak interaction model (\citealt{thomson90}, \citealt{thomson91}) is also able to produce long lasting ($\sim$10 Gyr), azimuthally distributed shell structures by the interference of density waves. In this scenario, the shell material does not originate from a companion galaxy, but comes from the thick disk population of dynamically cold stars of the galaxy. This is an unexpected result because galaxies that have a disk component should have been classified as S0s, rather than elliptical. Thus, it was difficult to explain the formation of shells in elliptical galaxies with a weak interaction model.
 But using two-dimensional stellar kinematics, SAURON (\citealt{bacon01}) and ATLAS$^{3D}$ (\citealt{cappellari11a}) studies have found that ETGs can be divided into two kinematically distinct families, slow and fast rotators (\citealt{emsellem07}; \citealt{emsellem11}). Fast rotators show large-scale rotation, while slow rotators show regular velocity field without large-scale rotation. These classifications are nearly insensitive to viewing angles. \citet{emsellem11} show that 66\% (45/68) of elliptical galaxies are fast rotators, which implies that they might have a disk component even though they were classified as ellipticals. Thus, at least in this fraction of elliptical galaxies, shells might be formed via weak interaction as well as via major or minor mergers.  
 
Van Dokkum (2005) reports that as many as 71\% (61/86) of bulge-dominated ETGs in the NOAO Deep Wide-Field Survey at $\langle  z \rangle$  $\sim$ 0.1 show morphological evidence of tidal interactions. A similar result was obtained by \citet{tal09}, who found that 73\% (40/55) of nearby (15 $-$ 50 Mpc)  elliptical galaxies show such features.  With deep V-band imaging data, \citet{janowiecki10} investigated the five brightest elliptical galaxies in the Virgo cluster and found that except for M84, all showed shells and diffuse tidal streams in their outer halos, in the range $\mu_{V}$ = 26 -- 29 mag arcsec$^{-2}$. At slightly higher redshifts, $\langle z \rangle \sim$ 0.15 -- 0.20, \citet{bennert08} studied the merger remnants of 5 early type QSO host galaxies and found tidal tails and shells in 4 of them.  These studies suggest that many elliptical galaxies have undergone recent mergers or interactions, consistent with \citeauthor{toomre77}'s \citeyearpar{toomre77} suggestion.  What remains less well-known is the amount of stellar mass in the tidal debris around ETGs. \citet{prieur88} investigated a shell galaxy, NGC 3923, using B-band images and found that the B-band luminosity of the shell contributes about 5\% of the galaxy luminosity. \citet{bennert08} also investigated the luminosity of shells in four early type quasi-stellar object host galaxies with HST V-band F606W images and estimated that shells comprises about 5 -- 10\% of the total V-band luminosity of each galaxy. The current study aims to measure the stellar mass fraction of the tidal debris using mid-infrared data at 3.6 {\micron} from the {\em{Spitzer}} Survey of Stellar Structure in Galaxies (\s4g, \citealt{sheth10}).

A second aim of this paper is to understand whether the recent merger or interaction, as revealed by the presence of tidal debris, has affected the overall structural properties of the galaxy itself. \cite{kormendy77} showed that elliptical galaxies and the bulges of S0 galaxies both have correlations between effective radius and surface brightness (Kormendy relation, hereafter KR). In addition, ETGs are known to follow empirical scaling relations between the effective radius, the surface brightness and the central velocity dispersion, referred to as the Fundamental Plane (hereafter FP,  \citealt{dressler87}; \citealt{djorgovski87}) or the Fundamental Manifold (\citealt{zaritsky06}; \citealt{zaritsky08}) if one extends FP formalism to lower or higher masses from dwarf spheroidal galaxies up to intracluster stellar populations of galaxy cluster.  
But not all ETGs lie on the FP; post starburst galaxies (E$+$A's) are one examples of galaxies that do not (\citealt{yang08}).  Similar to ETGs, E$+$A galaxies have large bulge to total luminosity ratios, large {\ser} indices, high concentration indices and a lack of on-going star formation, but have significantly larger asymmetry indices. 
However, these galaxies do not lie on the same FP with ETGs and follow their own relation, a tilted FP (\citealt{yang08}). \citet{yang08} showed that $55\%$ (11/21) of their E+A galaxies have dramatic tidal features, indicative of a recent merger/interaction. 
There might be two possible explanations for the offsets from the standard FP: changes in the structure, and/or changes in the underlying stellar populations.  
Through gas-poor (``dry'') merging events, galaxies can increase their half light radius (\citealt{nipoti03}; \citealt{boylan06}; \citealt{naab09}; \citealt{vanderwel09}; \citealt{hopkins09_dissipation_iv}) and their structural properties change and they may lie off from the FP of normal ETGs.
On the other hand, changes in the stellar populations due to star formation via gas-rich (``wet'') merging events may cause galaxies to have smaller mass to light ratios ($M/L$)
 and thus deviate from the standard FP.  For example, E+A galaxies are found to have 3.8 times smaller $M/L$ (\citealt{yang08}) and stand apart from the standard FP.
As tidal debris and shells around ETGs are indicative of recent merger or interaction, there might be changes in structure of those ETGs. Therefore we also address whether the ETGs with shells and tidal debris lie off the FP and/or KR.

The paper is organized as follows:  In Section 2, we describe the data and data analysis techniques.  
Then, we describe galaxies showing shells, tidal debris and the fraction of tidally disturbed galaxies in ETGs in Section 3.  
The luminosities of the tidal debris are presented and discussed in Section 4, and structural properties, KR and FP of tidally disturbed ETGs are 
explored in Section 5. Finally, we summarize our results and state our conclusions in Section 6.
 Throughout this paper, we adopt  $H_{0}=73$ \kms $Mpc^{-1}$, $\Omega_{m}=0.27$, and $\Omega_{\Lambda}=0.73$ when estimating distances.

\section{Observations and Data Analysis}
%\clearpage
\begin{deluxetable*}{lcccccr}
\tablecolumns{6}
\tablewidth{0pc}
\tabletypesize{\scriptsize}
\tablecaption{Catalog of parent sample of early type galaxies \label{tab_list_gal}}
\tablehead{ 
 \colhead{Object} &
 \colhead{T-type} &
 \colhead{Morphology} &
 \colhead{D25}&
 \colhead{$B_{T}$} &
 \colhead{$M_{B}$}&
 \colhead{$v_{rad}$}  \\
 \colhead{ } & & \colhead{Buta} & \colhead{[$\arcmin$]} & [Mag] &  [Mag]  & \colhead{ \kms] }  \\
 %\colhead{ } &  &\colhead{[$\arcmin$]} & [Mag] &  [Mag]  & \colhead{ [km/s] }  \\
 \colhead{(1)} &
 \colhead{(2)} &
 \colhead{(3)} &
 \colhead{(4)} &
 \colhead{(5)} &
 \colhead{(6)} &
 \colhead{(7)}
 \\
}
\startdata
   ESO 357$-$025 &     $-$2.8 &   \nodata  &  1.32 &     15.01 &      $-$16.75 &     1737  \\  
  ESO 358$-$025 &     $-$2.7 &   \nodata  &    1.74 &     13.83 &      $-$18.09 &     2184  \\  
  ESO 419$-$013 &     $-$1.0 &   \nodata  &    1.48 &     14.36 &      $-$17.00 &     1490  \\  
  ESO 462$-$031 &     $-$1.0 &   \nodata  &   1.62 &     14.62 &      $-$18.62 &     2705  \\  
  ESO 482$-$013 &     $-$1.0 &   \nodata  &   1.23 &     15.17 &      $-$16.80 &     1846  \\  
  ESO 483$-$013 &     $-$3.0 &   dE4,N/SA0$^{-}$  &   1.70 &     14.21 &      $-$16.04 &      895  \\  
  ESO 548$-$023 &     $-$3.7 &   \nodata  &   1.15 &     14.86 &      $-$17.25 &     1863  \\  
      IC 0051    &     $-$2.0 &     \nodata  &  1.70 &     13.75 &      $-$18.13 &     1714  \\  
      IC 2040    &     $-$1.1 &    \nodata  &   1.35 &     13.74 &      $-$17.38 &     1326  \\  
      IC 2085    &     $-$1.5 &    \nodata  &   2.45 &     13.96 &      $-$16.12 &      982  \\  
     NGC 0059 &     $-$3.0 &    \nodata  &   2.40 &     13.14 &      $-$15.17 &      364  \\  
     NGC 0148 &     $-$2.0 &    \nodata  &   2.00 &     13.11 &      $-$18.48 &     1516  \\  
     NGC 0216 &     $-$2.0 &    \nodata  &   1.35 &     13.71 &      $-$17.96 &     1576  \\  
     NGC 0244 &     $-$2.0 &    \nodata  &   1.05 &     13.79 &      $-$16.69 &      940  \\  
     NGC 0254 &     $-$1.3 &    \nodata  &   2.75 &     12.59 &      $-$19.11 &     1624  \\  
     NGC 0274 &     $-$2.8 &    (R)SA(l)0$^{-}$  &   1.26 &     13.40 &      $-$18.73 &     1751  \\  
     NGC 0474 &     $-$2.0 &    (R)SAB0/a (shells) pec  &   2.63 &     12.38 &      $-$20.43 &     2372  \\  
     NGC 0584 &     $-$4.6 &    S${\underline A}$B0$^{-}$   &   3.80 &     11.33 &      $-$20.89 &     1796  \\  
     NGC 0680 &     $-$4.0 &    \nodata  &   1.70 &     12.90 &      $-$20.52 &     2779  \\  
     NGC 0855 &     $-$4.9 &    SA0$^{-}$  &   2.95 &     13.28 &      $-$16.98 &      610  \\  
     NGC 0936 &     $-$1.2 &    \nodata  &   4.47 &     11.20 &      $-$20.27 &     1340  \\  
     NGC 1222 &     $-$3.0 &    \nodata  &   1.58 &     13.15 &      $-$19.80 &     2457  \\  
    NGC 1316C &     $-$2.0 &   \nodata  &   1.48 &     14.29 &      $-$17.61 &     1800  \\  
     NGC 1332 &     $-$2.9 &    \nodata  &   5.37 &     11.20 &      $-$20.46 &     1524  \\  
     NGC 1482 &     $-$0.9 &     Sa: sp    &   2.45 &     13.10 &      $-$19.02 &     1859  \\  
     NGC 1510 &     $-$1.6 &     SA0$^{+}$:  &   1.32 &     13.47 &      $-$16.49 &      913  \\  
     NGC 1533 &     $-$2.5 &    \nodata  &   3.24 &     11.79 &      $-$17.60 &      789  \\  
     NGC 1596 &     $-$2.0 &    \nodata  &    3.89 &     12.01 &      $-$19.28 &     1510  \\  
     NGC 1705 &     $-$2.9 &     dE3,N     &    1.86 &     12.79 &     \nodata &      630  \\  
     NGC 2634 &     $-$4.9 &    SA(nl)0$^{-}$ (shells)  &    1.70 &     12.91 &      $-$20.02 &     2258\tablenotemark{a}  \\  
     NGC 2732 &     $-$2.0 &    \nodata  &    1.82 &     12.85 &      $-$19.77 &     1900  \\  
     NGC 2768 &     $-$4.4 &    \nodata  &    5.62 &     10.82 &      $-$21.22 &     1410  \\  
     NGC 2974 &     $-$4.2 &     SA(r)0/a  &    3.47 &     11.88 &      $-$20.49 &     1889  \\  
     NGC 3032 &     $-$1.9 &    \nodata  &   1.41 &     13.07 &      $-$18.90 &     1546  \\  
     NGC 3073 &     $-$2.8 &     S0$^{-}$  &   1.20 &     14.11 &      $-$17.50 &     1217  \\  
     NGC 3384 &     $-$2.7 &    \nodata  &   5.25 &     10.89 &      $-$19.82 &      903  \\  
     NGC 3414 &     $-$2.0 &    \nodata  &   2.69 &     12.06 &      $-$19.80 &     1414  \\  
     NGC 3522 &     $-$4.9 &    \nodata  &   1.17 &     14.08 &      $-$17.41 &     1221  \\  
     NGC 3608 &     $-$4.8 &       E2        &   3.16 &     11.57 &      $-$19.74 &     1108  \\  
     NGC 3773 &     $-$2.0 &    \nodata  &   1.23 &     13.51 &      $-$17.52 &      985  \\  
     NGC 3870 &     $-$2.0 &    SB(rs)0$^{o}$?  & 1.02 & 13.49 &   $-$17.32 &    758  \\  
     NGC 4105 &     $-$4.6 &    \nodata  &    4.37  &   11.57  &     $-$20.72 &   1873 \\
     NGC 4106 &     $-$1.3 &   \nodata  &    4.17 &    12.28  &      $-$20.34 &    2150  \\  
     NGC 4117 &     $-$2.0 &     S0$^{-}$ sp  &  1.58 &     14.10 &      $-$17.04 &      934  \\  
     NGC 4405 &     $-$0.1 &    \nodata  &   1.70 &     12.96 &      $-$19.35 &     1741  \\  
     NGC 4494 &     $-$4.8 &    \nodata  &    4.37 &     10.68 &      $-$21.02 &     1310  \\  
     NGC 4546 &     $-$2.7 &    \nodata  &   3.16 &     11.35 &      $-$19.73 &     1050  \\  
     NGC 4550 &     $-$2.1 &    S0$^{-}$ sp  &   3.16 &     12.48 &      $-$18.69 &      381  \\
     NGC 4694 &     $-$2.0 &    \nodata  &   2.00 &     12.29 &      $-$19.19 &     1182  \\
     NGC 5018 &     $-$4.4 &    SAB0$^{-}$ (shells) pec  &   3.47 &     11.69 &      $-$21.64 &     2850  \\
     NGC 5122 &     $-$1.0 &    \nodata  &   1.15 &     14.27 &      $-$18.98 &     2859  \\
     NGC 5173 &     $-$4.9 &    E$^{+}$  &   1.10 &     13.38 &      $-$19.68 &     2428  \\
     NGC 5273 &     $-$1.9 &    \nodata  &   2.29 &     12.53 &      $-$18.86 &     1109  \\
     NGC 5338 &     $-$2.0 &    \nodata  &   1.95 &     14.05 &      $-$16.61 &      816  \\
     NGC 5481 &     $-$3.9 &    \nodata  &   1.78 &     13.28 &      $-$19.26 &     1883  \\
     NGC 5574 &     $-$2.8 &    \nodata  &   1.32 &     13.29 &      $-$18.79 &     1657  \\
     NGC 5576 &     $-$4.8 &    \nodata  &   2.82 &     11.79 &      $-$20.14 &     1482  \\
     NGC 5846 &     $-$4.7 &    E$^{+}$0/SA0$^{-}$  &   4.27 &     11.09 &      $-$21.27 &     1824  \\
     NGC 5854 &     $-$1.1 &    \nodata  &   3.02 &     12.65 &      $-$19.70 &     1750  \\
     NGC 5864 &     $-$1.7 &    \nodata  &   2.51 &     12.68 &      $-$19.76 &     1850  \\
     NGC 6014 &     $-$1.9 &    \nodata  &   1.74 &     13.64 &      $-$19.46 &     2491  \\
     NGC 6278 &     $-$1.9 &    \nodata  &   1.74 &     13.74 &      $-$19.73 &     2776  \\
     NGC 7465 &     $-$1.9 &    \nodata  &   1.07 &     13.36 &      $-$19.33 &     1962  \\
  PGC 014037 &     $-$1.0 &    \nodata  &   1.20 &     15.12 &      $-$17.74 &     2546 \\
    UGC 09348 &     $-$1.6 &    \nodata  &   1.78 &     14.72 &      $-$17.45 &     1672 
\enddata
\tablecomments{ Information on galaxies checked for tidal debris. All data were taken from Hyperleda, unless noted otherwise.
Col. (1) : Object.
Col. (2) : Numerical morphological type.
Col. (3) : Galaxy morphology at $3.6${\micron} (\citealt{buta10}).
Col. (4) : Projected major axis of a galaxy at the isophotal level of $B=25$ $mag/arcsec^{2}$.
Col. (5) : Apparent total B magnitude.
Col. (6) : Absolute B magnitude.
Col. (7) : Heliocentric radial velocity obtained from radio observations.}
\tablenotetext{a}{: Heliocentric velocity from \citet{smith00}. This galaxy does not have a value for $v_{rad}$ in the Hyperleda database at the time of writing, hence it is not included in the \s4g sample. But as the nearby galaxy, NGC 2634A is in the \s4g sample, NGC 2634 was imaged in the same frame.
}
\end{deluxetable*}

\subsection{Data}
 \s4g is one of the ten large Exploratory Science programs underway in the {\em{Spitzer}} post-cryogenic mission. The overall goal of \s4g is to image over 2,300 nearby galaxies at 3.6 and 4.5 {\micron} through new ``Warm" observations  obtained with the Infrared Array Camera (IRAC, \citealt{fazio04}) instrument on the {\em{Spitzer}} Space Telescope (\citealt{werner04}). These data will add to the existing data of galaxies from the {\em{Spitzer}} Heritage Archive. \s4g sample galaxies have radial velocity $V_{radio} <$ 3,000 {\kms} corresponding to d $<$ 40 Mpc, a total corrected blue magnitude, $m_{Bcorr} <$ 15.5 mag, and blue light isophotal angular diameter $D_{25} > 1\arcmin$,  while $ \left| b \right| > 30\arcdeg$.  The \s4g data are probing down to stellar surface densities $<$ 1 M$_{\sun}$ pc$^{-2}$.  The azimuthally averaged surface brightness profiles typically trace isophotes down to $\mu_{3.6 {\: \mu} m} >$ 27 (AB) mag $arcsec^{-2}$. All of the images are processed using the \s4g pipeline (see details in \citealt{sheth10}). The resulting pixel scale of \s4g image is 0.75$\arcsec$/pixel and the typical full width half maximum of the Point Spread Function (PSF) is 1.7$\arcsec$. 
Using the zero magnitude flux at 3.6 {\micron}, 280.9 Jy, as indicated in the \textit{IRAC Instrumental Handbook}, 3.6 {\micron} AB magnitude can be converted from 3.6 {\micron} Vega magnitude using the following equation
  \begin{equation}
M_{3.6, AB} = M_{3.6, Vega} + 2.78
\end{equation}
 
To identify tidal features in ETGs, we first chose ETGs from \s4g using their numerical Hubble type T (\citealt{rc3}), which was obtained from Hyperleda (\citealt{paturel03}). We selected ETGs with T less than or equal to $0$, thus our sample includes both Ellipticals and S0s. Note that although \s4g surveys volume-, magnitude-, and size-limited samples, there is a caveat especially on the completeness of ETG samples.  Since \s4g chose galaxies with radial velocity $V_{HI} < 3,000$ {\kms}  for sample selection, \s4g misses gas-poor faint and small ETGs. 
We draw and analyze an initial set of 65 ETGs that were available at the moment of this pilot study, of which 34 are from the cryogenic mission archival data and 31 are newly obtained during the post-cryogenic phase of {\em{Spitzer}}. The basic properties of the 65 ETGs are listed in Table \ref{tab_list_gal}. Our sample consists of 16 ellipticals (T $<-3.5$; criteria adopted from \citealt{paturel03} and \citealt{emsellem11}) and 49 S0s ($-3.5 \geq $ T $\le 0$). There will be 180 ETGs in the complete sample of 2,300 \s4g galaxies. ETGs with tidal debris presented in this study are located at distance between 20 and 43 Mpc, so the typical PSF (1.7$\arcsec$ ) corresponds to a linear scale of 170 to 350 pc.

\subsection{Data Analysis}
The principal difficulty in identifying tidal debris is the very high contrast between the bright component of a galaxy and the faint, fine-scale tidal debris structures. To mitigate this contrast, we used the following methods: unsharp masking (\citealt{malin77}), the building of a structure map (\citealt{pogge02}) and color map, and the two-dimensional profile fitting of the light distribution of the galaxy. Applying unsharp masking is a fast and easy way to remove the light associated with the galaxy. It enhances the features of galaxies that exhibit sudden changes in their light distributions, which can be caused by spiral arms, shells and ripples. To generate an unsharp masked image, we first convolved the image with a circular Gaussian kernel of $\sigma = 15$ pixels (corresponding to $11.25\arcsec$), a size that was chosen iteratively to better reveal the tidal debris in the unsharp masked images.  We then divided the original image by the convolved image, which allowed us to highlight the sharply varying features from the smoother background.  This technique was sufficient for revealing shells, arc-like features (see Figure \ref{ngc2974_usm}), and highly-inclined disk components.

Making structure maps (\citealt{pogge02}) is another method to  enhance unresolved or marginally resolved features on the scale of the PSF. This technique involves removing the large scale smoothed light of the galaxy using the image PSF as a kernel. The structure map therefore has the extra benefit of reducing the time spent in choosing the best kernel to reveal the tidal features of the galaxy.
To derive the structure map, we first divide the original image by the PSF-convolved image, then convolve this resulting image with the transpose of the PSF image. This operation can be expressed as follows, 
\begin{equation}
S = \left[ \frac{I}{I \otimes P} \right] \otimes P^{t},
\end{equation}
where $I$ is the original image, $P$ is the PSF image, $P^{t}$ is the transpose of PSF image, and $S$ is the image of the structure map  (\citealt{pogge02}).  Smooth changes in the light distribution, such as faint tidal tails, are still not easily revealed with an unsharp masked image or structure map. 

Modeling the light distribution of the galaxy using its known components such as a bulge and/or a disk component is another common method, used for detecting tidal features. The {\sc{Ellipse}} task in {\sc{iraf}} can be used for modeling the light distribution of the galaxy. This task creates series of concentric elliptical isophotes that are fit to the galaxy image. The resulting galaxy model is subtracted from the original image to reveal the tidal features. This technique is, however, not optimal to identify shells because concentric arc-like shell features can become incorporated in the {\sc{Ellipse}} model. Another problem with the ellipse-fitting method is that it can produce spurious residuals when there are rapid changes in the ellipticity and position angle (PA).  In such cases, the model subtraction leaves an imprint, which resembles shell-like debris (\citealt{janowiecki10}).   
We also found in our analysis that the residual images of the {\sc{Ellipse}} model fit can show  ``butterfly'' (quadrupole) patterns in the central region (also in \citealt{janowiecki10}). This occurs because the ellipse fitting cannot account for high order deviations from an ellipses (e.g., boxy bulges), which may be present near galactic centers.  

\setlength{\tabcolsep}{2.0pt}
\begin{deluxetable*}{cccccccrrc}
\tablecolumns{10}
\tabletypesize{\scriptsize}
\tablewidth{0pc}
\tablecaption{Basic Information of the Galaxies showing tidal features. \label{tab_basicinfo}}
\tablehead{ 
  \colhead{Object} &
  \colhead{Morphology} &
  \colhead{Morphology} &
  \colhead{Kinematic class }&
    \colhead{$\Psi$ }&
  \colhead{Distance } &
  \colhead{$B_{Total}$} &
  \multicolumn{3}{c}{Velocity dispersion}\\
  \colhead{} & \colhead {(RC3)} & \colhead {(Buta)} &\colhead {($ATLAS^{3D}$)} & \colhead {[deg]} & \colhead {[Mpc]} & \colhead{[Mag]} &  \multicolumn{3}{c}{[\kms]} \\
  \colhead{(1)} & \colhead{(2)} & \colhead{(3)} & \colhead{(4)} & \colhead{(5)} & \colhead{(6)} & \colhead{(7)} & \multicolumn{3}{c}{(8)} } 
\startdata
NGC 0474 &    SA(s)$0^{o}$      &   (R)SAB0/a (shells) pec     &  Fast Rotator  &     31.3      & 27.7   &  12.38     &    163  &$\pm$& 5$^{(a)}$ \\ 
NGC 0680 &    E$^{+}$ pec$:$  &   \nodata                          &  Fast Rotator  &     22.7      & 34.8   &  12.90     &    214  &$\pm$& 6$^{(a)}$ \\
NGC 1222 &    S0$^{-}$ pec$:$ &  \nodata                           &  Slow Rotator  &    72.7     &  31.2   & 13.15     &     119  &$\pm$&13$^{(b)}$ \\
NGC 1482 &    SA0$^{+}$ pec sp   & Sa$:$ sp                        &  \nodata        &   \nodata  & 25.0   & 13.10    & &\nodata  &\\
NGC 2634 &    E1$:$                 &   SA(nl)0$^{-}$ (shells)      &  \nodata        &   \nodata   & 31.7  &  12.91   &     191  &$\pm$&6$^{(a)}$ \\ 
NGC 3032 &    SAB(r)0$^{0}$     & \nodata                             &  Fast Rotator  &    0.8        &  25.0  & 13.07    &     82   &$\pm$&16$^{(c)}$ \\
NGC 4106 &    SB(s)0$^{+}$      & \nodata                             &  \nodata        &   \nodata   & 34.3  &  12.28    &  176    &$\pm$&17$^{(d)}$ \\
NGC 4694 &    SB0 pec             &   \nodata                            &  Fast Rotator  &    2.0         & 20.4  &  12.29   &     61    &$\pm$&5$^{(b)}$ \\ 
NGC 5018 &    E3$:$                &   SAB0$^{-}$ (shells) pec    &  \nodata         &   \nodata & 43.3  &   11.69   &   211    &$\pm$&17$^{(b)}$ \\                                                    
NGC 5574 &    SB0$^{-}:?$ sp  &   \nodata                             &  Fast Rotator  &   4.8         & 26.3  &   13.29   &    77     &$\pm$&3$^{(f)}$ \\ 
NGC 5576 &    E3                     &   \nodata                            &  Slow Rotator  &   7.4        & 23.9  &   11.79   &   164     &$\pm$&13$^{(b)}$
\enddata
\tablecomments{ 
Col. (1): Object.
Col. (2): Galaxy morphology in the optical blue band from RC3 catalog.
Col. (3): Galaxy morphology at $3.6${\micron} (\citealt{buta10}).
Col. (4): Kinematic class of galaxies from $ATLAS^{3D}$ study (\citealt{emsellem11}).
Col. (5): Kinematic misalignment angle in degrees from $ATLAS^{3D}$ study (\citealt{krajnovic11}), where sin $\Psi = \mid$ sin(PA$_{phot} -$ PA$_{kin}) \mid$.
Col. (6): Luminosity distances which were corrected to the reference frame defined by the 3K microwave background radiation, from the NASA/IPAC Extragalactic database (NED).
Col. (7): Total B magnitude, from Hyperleda.
Col. (8): Central velocity dispersion, from (a): \citet{simien00}; (b): Table 4 of \citet{wegner03}; (c): \citet{dalleore91}; (d): \citet{dressler91}  (f): \citet{simien02}}
\end{deluxetable*}

A better technique for identifying tidal features is to use a multi-component, two-dimensional fitting algorithm like {\sc{GALFIT}} (version 3.0, \citealt{peng10}; \citealt{peng02}). 
In order to find tidal debris, we examine the residual image, which were obtained by subtracting the best fit model from the image. 
To find the best fit model for each galaxy, we run {\sc{GALFIT}} with four different functions; two different one-component fits that fit the bulge only and two different two-component fits that fit the bulge and the disk. These four functions are de Vaucouleurs, {\ser}, {\ser}+exponential and {\ser}+{\ser} profile. These fits will be discussed in more detail in Section 5.1.
Although it has been shown that most of the ETGs do not follow a $R^{1/4}$ law (\citealt{caon93}; \citealt{graham96}), i.e., the {\ser} index of the bulge is rarely 4, we modeled the ETGs with a n$=$4, de Vaucouleurs' profile to compare our results with previous studies on the KR and FP. Since some elliptical galaxies are mistaken S0s, we include a disk component for modeling those galaxies to increase our chances of diagnosing such cases. If we choose an unrealistic {\ser} index to model galaxies, spurious features can also be produced in the residual image, therefore careful decomposition is needed.

The {\ser} profile (\citealt{sersic63}) is defined as, 
\begin{equation}
I(r) = I_{\rm eff}~{\rm exp} \left\{ -b_n \left[ \left( \frac{r}{r_{\rm eff}}\right)^{1/n}-1 \right] \right\},
\end{equation}
\noindent
where $r_{\rm eff}$ is the effective radius, $I_{\rm eff}$ is the intensity at $r=r_{\rm eff}$,
$n$ is the {\ser} index, and $b_n$ is chosen to satisfy
\begin{equation}
\int_0^{\infty}I(r) 2 \pi r dr = 2 \int_0^{r_{\rm eff}} I(r) 2 \pi r dr.
\end{equation}
When $n=4$, it is called the de Vaucouleurs profile, and if $n=1$, it describes the exponential disk profile.
The scale length $h_{r}$ of the disk is given as,
\begin{equation}
r_{\rm eff} = 1.678 h_{r}.
\end{equation}

{\sc{GALFIT}} gives the effective semi major axis in the output, whereas other studies often use, the circularized effective radius, defined as 
\begin{equation}\label{circular_reff}
R_{\rm eff}=r_{\rm eff} \times \sqrt{b \over a}.
\end{equation}
We also derive $R_{\rm eff}$ for comparison with other studies.
Mean surface brightnesses within $r=r_{\rm eff}$ are calculated with
\begin{equation}
<\mu_{\rm eff}> = m_{\rm bulge} + 2.5 log [ \pi {R_{\rm eff}}^2] ,
\end{equation}
where $m_{\rm bulge}$ is the bulge magnitude within the effective radius, $(b/a)$ is the axis ratio of the bulge component, and $r_{\rm eff}$ is effective radius of the bulge.

 We constrained {\sc{GALFIT}} to use the same central position of the galaxy for both the bulge and disk components.
{\sc{GALFIT}} (version 3.0 or higher) allows us to fit asymmetric features with Fourier modes and coordinate rotations. Unlike {\sc{Ellipse}} in {\sc{iraf}}, {\sc{GALFIT}} does not build models using isophotal fits, but instead fits the given image with a suitable model all at once.
Thus, it does not allow radial variations of PA and ellipticity but returns one global value of PA and ellipticity for each component of the model.
 We estimated the sky background by selecting 10 different regions outside the galaxy, avoiding any obvious faint tidal features in the background. We fixed the sky value while running {\sc{GALFIT}}. All of the components were convolved with the PSF for IRAC 3.6 {\micron}-band, which was created from the sub-sampled IRAC Point Response Function (PRF) provided by the {\em{Spitzer}} Science Center\footnote{http://ssc.spitzer.caltech.edu/irac/calibrationfiles/psfprf}. 
Because images of the {\s4g} are mosaicked by drizzling, and have a pixel scale of 0.75$\arcsec$, we re-sampled the PRF 61 times (=0.75/1.22$ \times$100) to match the pixel scale of the \s4g data. We also rotated each PSF to the average angle at which a galaxy was observed. 
For the Galactic extinction correction, we use the result of  \cite{schlegel98}. Because the difference of the Galactic extinction between 3.6 and 3.8 {\micron} is not significant, we adopted the value of $L^{\prime}$ (3.8 {\micron}) for each object, taken from the NASA/IPAC Extragalactic Database (NED)\footnote{http://ned.ipac.caltech.edu}.

To find the best fit model unaffected by bright foreground stars or faint background galaxies, we mask out those objects before running {\sc{GALFIT}}.
To generate mask images, we run Source Extractor (\citealt{bertin96}) with a 3 sigma detection threshold and choose
``check-image'' to return ``segmentation image'' and make use of those as mask images. However, the ``segmentation image'' does not generate adequate
masks, especially in the central parts of some galaxies. So for sources lying within 0.5$\times D_{25}$ from the center of
galaxies, we create circular masks with the size proportional to the magnitude of the source. Then, we perform a preliminary {\sc{GALFIT}} run to identify tidal
features. 
Based on this first analysis, we improve the mask images manually by adding shells or diffuse tidal features to the masks, which do not originally include those features in the ``segmentation image''.  
Then, we run {\sc{GALFIT}} again on this star-free, background galaxy-free, and tidal feature-free image to find the best fit.
In some cases, around the edge of masks, light from stars appears to be leaking out as can be seen in Figure 1 to 5.
However, these images are displayed with a histogram equalized stretch that emphasizes faint structures and the leaks do not affect our results.

%--------------------------------------
\begin{figure*}
\begin{center}
\includegraphics[width=16cm]{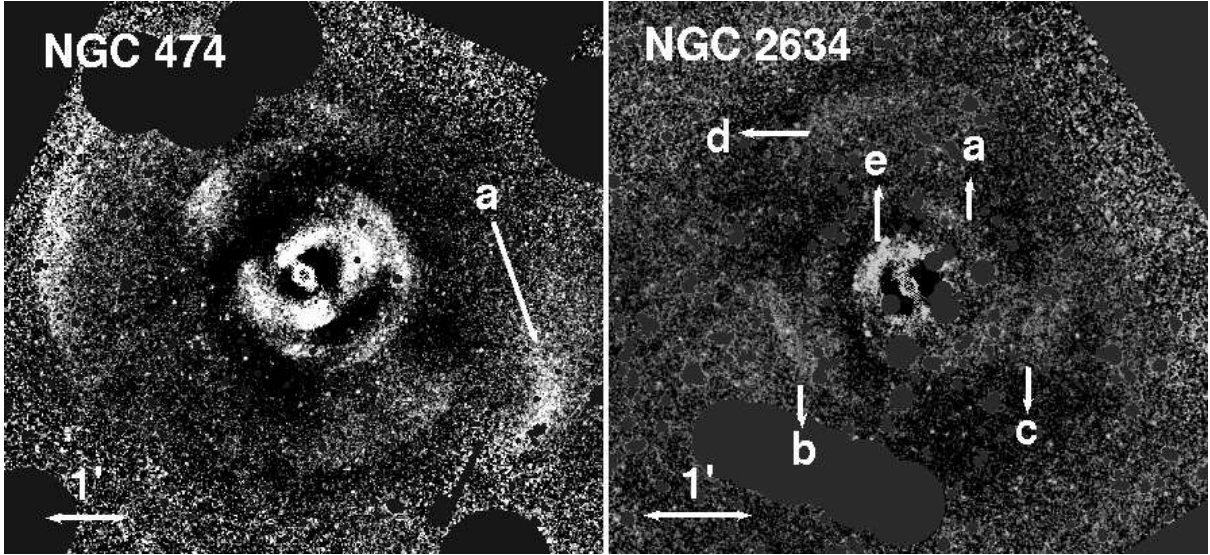}
\caption{ Residual images of NGC 474 (left) and NGC 2634 (right). The model fit of two {\ser} components, one for the bulge and one for the disk, is subtracted from each image. Left: Shells and tidal features of NGC 474. The``comma-like'' feature is marked with `a' at r$\sim$3.5$\arcmin$. Right: Shells are marked with `a' (r$\sim$45$\arcsec$), `b',`c' (r$\sim$75$\arcsec$), `d' (r$\sim$100$\arcsec$), and a possible lens is marked with `e' (r$\sim$25$\arcsec$). 
North is up and East is to the left for all of the images shown in this paper.
} \label{ngc0474_2634}
\end{center}
\end{figure*}

\section{Galaxies exhibiting tidal features}
Of the 65 ETGs in our sample, 11 galaxies show evidence of tidal interactions, such as shells, ripples or tails in the {\sc{GALFIT}} residual image. We also check the unsharp masked image, structure map. Especially for the diffuse features, we visually cross-check those features in the 3.6 {\micron} image using the histogram stretch and in the {\sc{GALFIT}} residual images. The basic properties of each galaxy are summarized in Table \ref{tab_basicinfo}. We describe properties of galaxies showing shells and tidal debris in Section 3.1 and 3.2, respectively. All of 11 galaxies were previously classified as galaxies with shells or tidal features, therefore we briefly summarize published studies on them. Then we present the fraction of ETGs with shells and tidal disturbances and we compare them with literatures in Section 3.3.  We also discuss the ring-like feature in NGC 2974, which may not be tidal debris.

\subsection{Shell galaxies}
$\bullet$ NGC 474:  This galaxy is also known as Arp 227 (\citealt{arp66}), and is interacting with NGC 470, which is 5.5$\arcmin$ ($\sim 50$ kpc) to the West.  An HI gas bridge connects NGC 474 and NGC 470 (\citealt{rampazzo06}). NGC 474 is famous for its shell structures (\citealt{malin83}; \citealt{turnbull99}) and the misalignment between the kinematic and photometric major axis (\citealt{hau96}).  About 3.5$\arcmin$ to the South-west of the galaxy center, there is a ``comma-like'' structure (\citealt{rampazzo06}), which is denoted with a letter `a' in Figure \ref{ngc0474_2634}. 

\begin{figure*}
\begin{center}
\includegraphics[width=16cm]{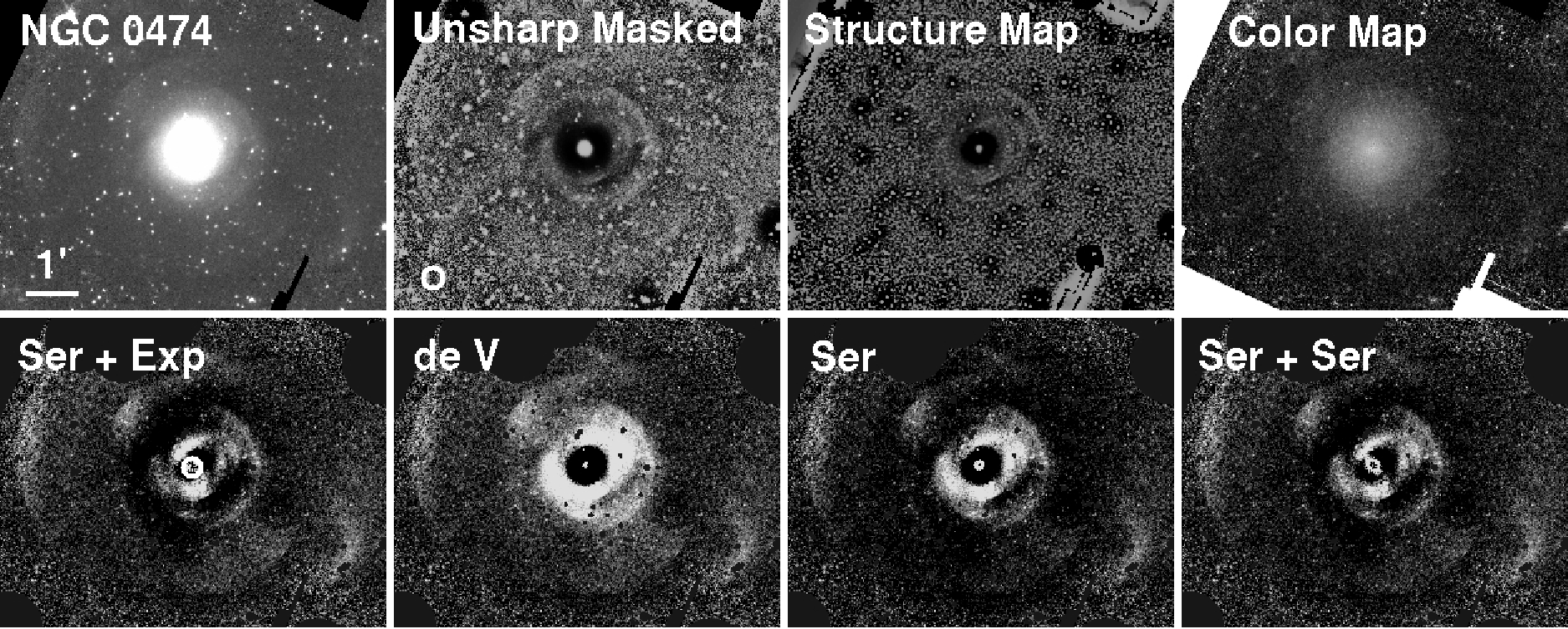}\\
\vspace{2 mm}
\includegraphics[width=16cm]{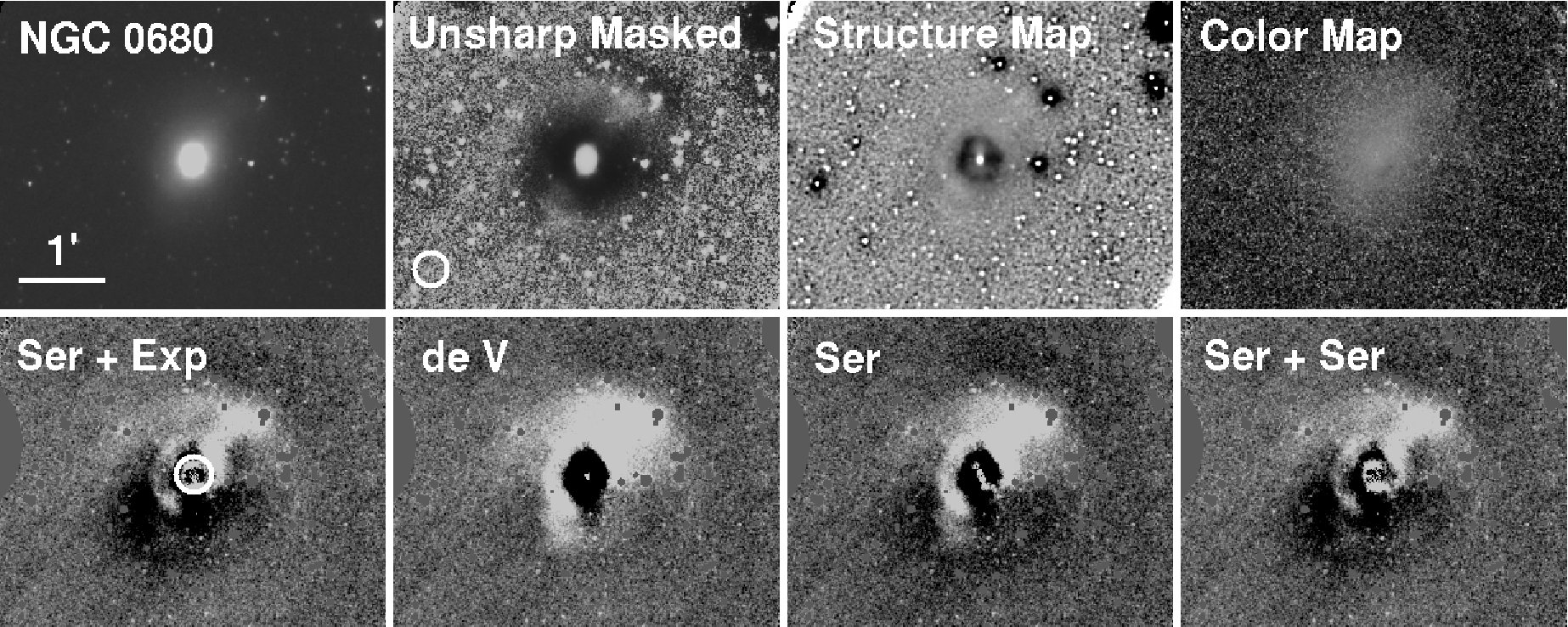}
\caption{ Image analysis of NGC 474 and NGC 680.
Top(\textit{left to right}): $3.6${\micron} image; unsharp masked image; structure map; and color image of $3.6 - 4.5$ {\micron} in histogram equalized stretch.
Bottom (\textit{left to right}): {\sc{GALFIT}} residual images for a bulge (n=free) and an exponential disk (n=1) fit; de Vauouleurs law (n=4) fit; bulge-only (n=free) fit, and bulge (n=free) with disk (n=free) fit, respectively. The white circle on the bottom left of the unsharp masked image indicates the size of the Gaussian kernel, which was used to generate the unsharp masked image. The white circle in the residual image of the `Ser+Exp' model presents the adopted inner boundary for estimating the fractional luminosity of tidal debris, which is also listed in Table \ref{tab_ratio_result}.}
\label{ngc0474set6} \label{ngc0680set6} 
\end{center}
\end{figure*}

In the \s4g data, shells in NGC 474 are clearly visible in the unsharp masked image and in the residual image (Figure \ref{ngc0474set6}) from the {\sc{GALFIT}} model. There is also a possible bar feature (PA 33$\arcdeg$ from North to East), which shows up in the residual image after a {\ser} plus the exponential disk model has been subtracted (Figure \ref{ngc0474_2634}). This is consistent with the findings of \cite{laurikainen10}, who also noted that this galaxy has a weak oval bar surrounded by a lens. 
A lens is defined as a oval-shaped structure in the disk and its surface brightness is slowly decreasing with a sharp outer edge (\citealt{kormendy79}; \citealt{athanassoula83}; \citealt{bosma83}; \citealt{kormendy04}). 
We also find that shells and tidal features in the residual images at 3.6 {\micron} are similar to those in R-band images obtained by \citet{turnbull99}. 
The PA and ellipticity of NGC 474 are in agreement with what \citet{turnbull99} have found. The 3.6$-$4.5 {\micron} color map is shown in Figure \ref{ngc0474set6}. 
Morphological features of NGC 474 seen in the 3.6$-$4.5 {\micron} color map are similar to those in the 3.6 {\micron} image,  and this is same as for other galaxies showing tidal features in this study.

%--------------------------------------
$\bullet$ NGC 2634: This galaxy presents multiple shells at r $\sim$  45, 75 and 100$\arcsec$, which correspond to projected distances of r $\sim$ 7 kpc (marked with `a' in Figure \ref{ngc0474_2634}), 11 kpc (`b' and `c'), and 15 kpc (`d'). In the residual images (Figure \ref{ngc2634set6}), a relatively bright ring-like feature is present at r$\sim$25$\arcsec$ as marked with `e'.  It is difficult to decide whether this is a shell or not because it does not show a sharp edge on the surface brightness profile and it encircles the central part of the galaxy. Therefore, it is more probable that the ring-like feature is a lens, as \cite{buta10} classified this galaxy has a nuclear lens. There are only a few previous studies that have noted these shells. While the galaxy was originally classified as ``E1:'' in the RC3,  it was recently classified by \cite{buta10} as a $SA(nl)0^{-}$ (shells) from the \s4g data. In a conference abstract, \cite{statler01} report that they found shells in NGC 2634 using optical image.

$\bullet$ NGC 3032: The galaxy has shell at r$\sim$30$\arcsec$ and it is marked with arrows in Figure \ref{NGC3032_set4}. In the residual image, we see some features in the central region. We check the $F606W$ $HST$ and Sloan Digital Sky Survey (SDSS) image and found that NGC 3032 has central dusty spiral features within 11$\arcsec$. \citet{schweizer88} also showed that this galaxies have 3 interleaving ripples in Figure 2 of their paper. \citet{laurikainen10} noted that this galaxy has two nearly circular lenses extending out to r$\sim$7 and 16$\arcsec$, respectively.

%--------------------------------------
$\bullet$ NGC 5018: NGC 5018 shows multiple shells (\citealt{malin83}) and broad fans of stellar light (Figure \ref{ngc5018set6}). One of the inner shells is in the shape of an hourglass, which is also evident in the V-band image in Figure 2 of \citet{fort86}. Dust lanes and patches in the galaxy have been noted (\citealt{fort86}; \citealt{sandage94}) and with a UV $F336W$  $HST$ image, \citet{rampazzo07} find dust lanes also in the nuclear region (10$\arcsec \times 10\arcsec$) of the galaxy. A filament of HI gas stretches across the stellar body of NGC 5018 and connects two neighbors (\citealt{serra10}).
 A faint dispersed tidal tail lies toward the North-west. The shells are not detected in a near-ultraviolet image taken from the \textit{GALEX} archive (\citealt{rampazzo07}), indicating that there is no strong recent star formation in shells.  The unsharp masked image, structure map, and residual images (Figure \ref{ngc5018set6}) suggest that there is an inner disk inclined at a PA of 97$\arcdeg$ in the central region of the galaxy, extending out to 25$\arcsec$ from the galactic center.  

%--------------------------------------
\subsection{Galaxies showing tidal features }
$\bullet$ NGC 680:  This galaxy is classified as a ``E$^{+}  $ pec$:$'' (RC3), and it belongs to the NGC 691 group, (\citealt{vanmoorsel88}; \citealt{crook07}) with NGC 678 (at a projected distance of 5.4$\arcmin$, 60 kpc ), NGC 694 (16.6$\arcmin$, 190 kpc), NGC 691 (17.9$\arcmin$, 200 kpc, also known as NGC 674),  and IC 167 (19.1$\arcmin$, 220 kpc). In our analysis of the \s4g data, we clearly see tidal disturbances, shells, and an asymmetric envelope in the residual maps from GALFIT shown in Figure \ref{ngc0680set6}. 
As an exercise, we have also run the {\sc{ellipse}} task in {\sc{iraf}} with a varying center to check if we can confirm the presence of the asymmetry in NGC 680. With this fit we can reproduce the asymmetric envelope.
However, the model image produced from the {\sc{ellipse}} fit shows artificial structures caused by the pile up of ellipses. Using this model to produce residual image similarly as with our {\sc{GALFIT}} models, we still find the
same tidal features, although their shape might vary. The dark feature
at the South-East side of NGC 680 in Figure \ref{ngc0680set6} becomes smaller in size -- and part of it becomes bright -- but the bright feature at the North-West, which we describe as a shell, still remains. NGC 680 has also been described to have possible patchy dust (\citealt{ebneter88}). \cite{duc11} showed that NGC 680 exhibits two diffuse plumes in the optical image and these are connected to an extended HI tail. 

%-------------------------------------------------------
\begin{figure*}
\begin{center}
\includegraphics[width=16cm]{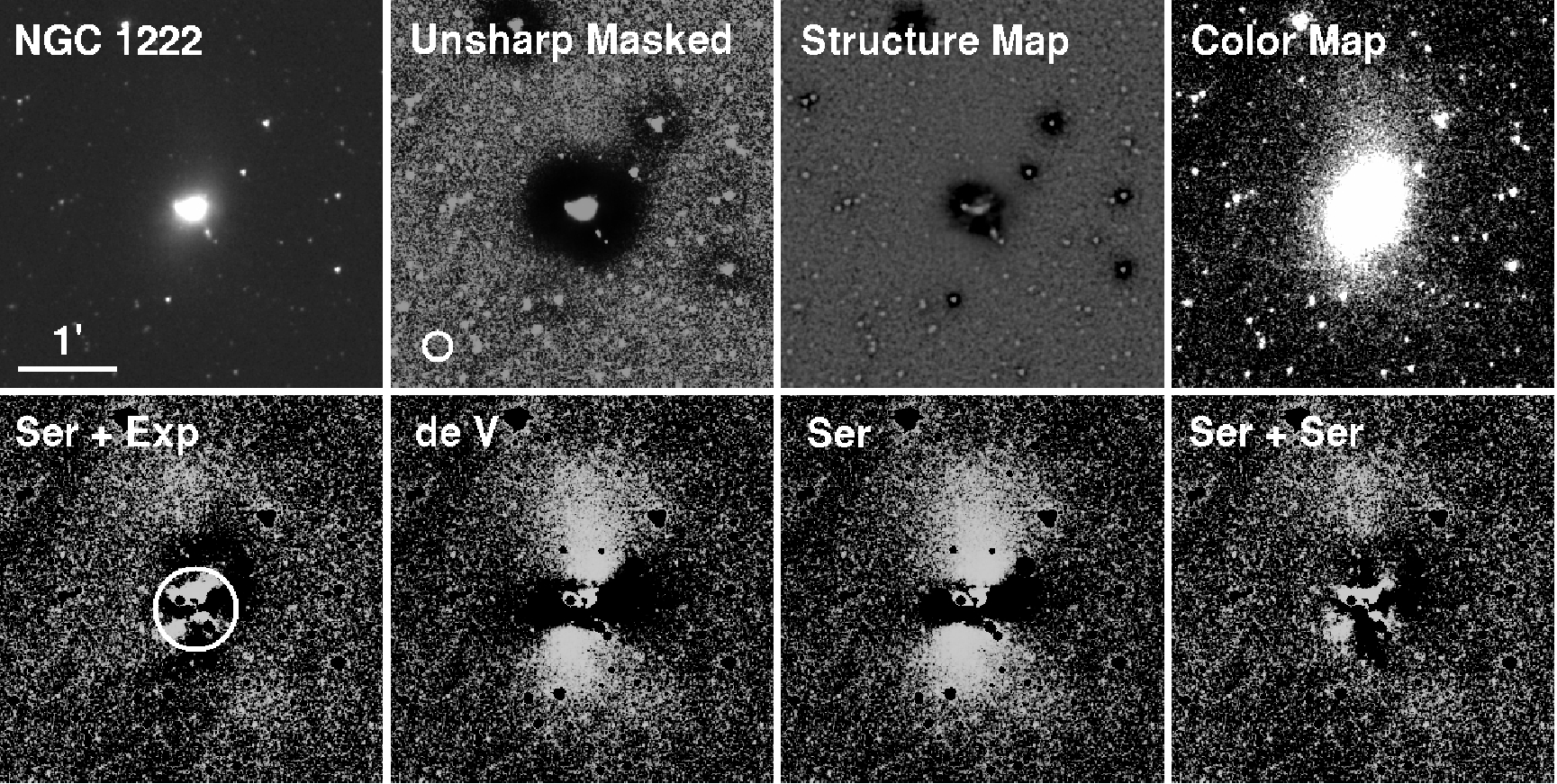}\\
\vspace{2 mm}
\includegraphics[width=16cm]{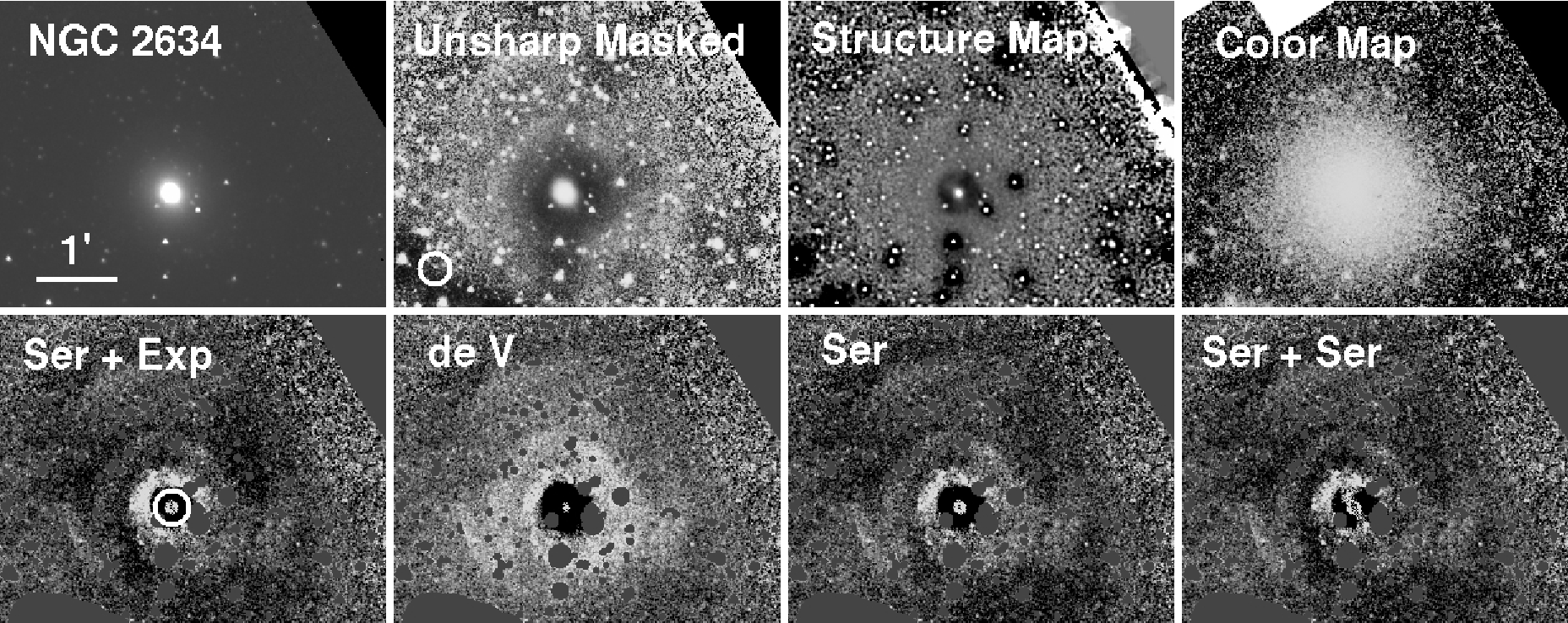}\\
\vspace{2 mm}
\includegraphics[width=16cm]{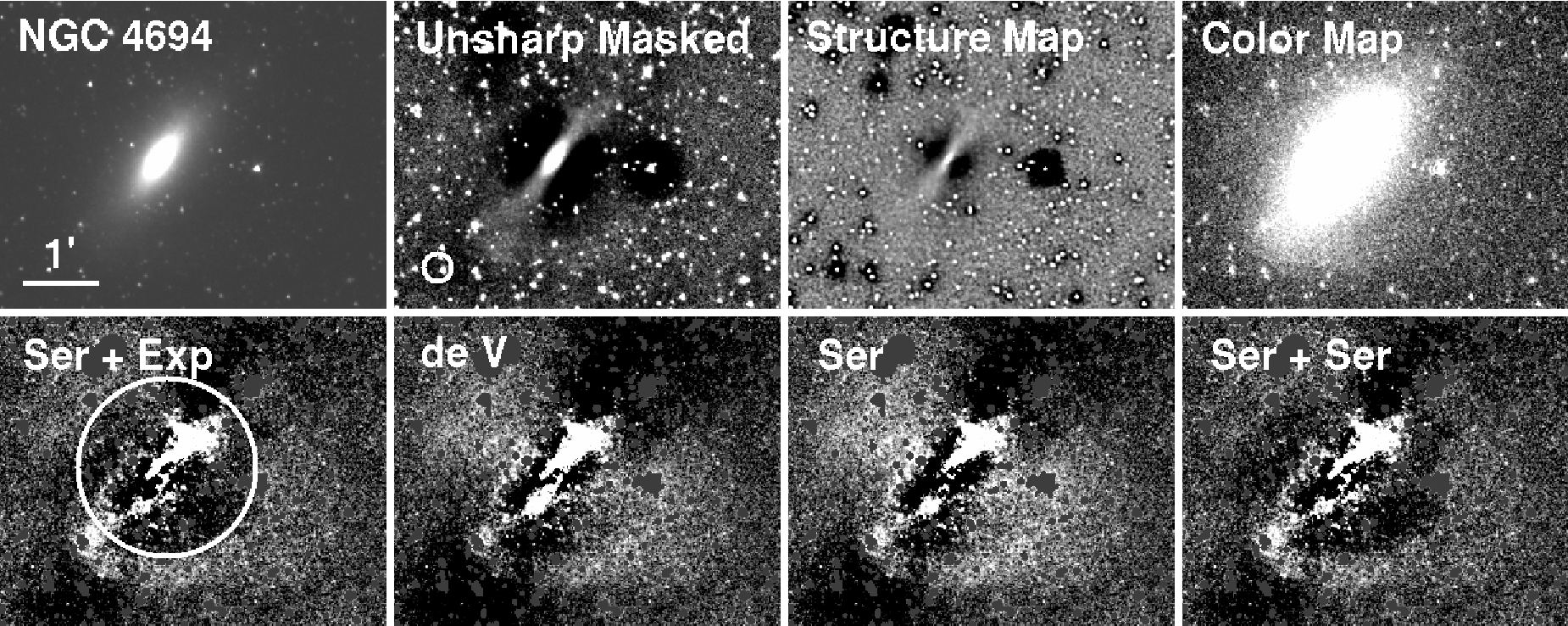}\\
\caption{ Image analysis of NGC 1222, NGC 2634 and NGC 4694, same as figure \ref{ngc0474set6}.
\label{ngc2634set6}  \label{ngc4694set6} \label{ngc1222set6} }
\end{center}
\end{figure*}
%-------------------------------------------------------
\begin{figure*}
\begin{center}
\includegraphics[width=16cm]{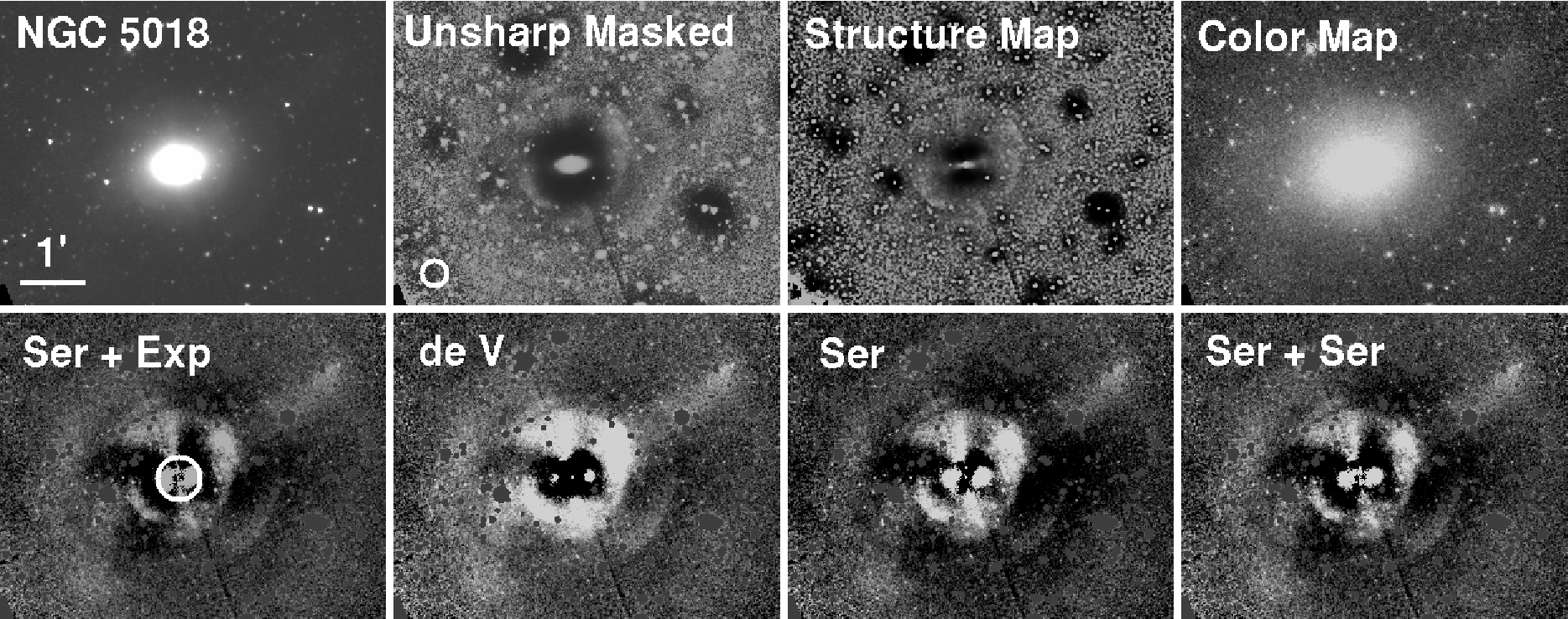}\\
\vspace{2 mm}
\includegraphics[width=16cm]{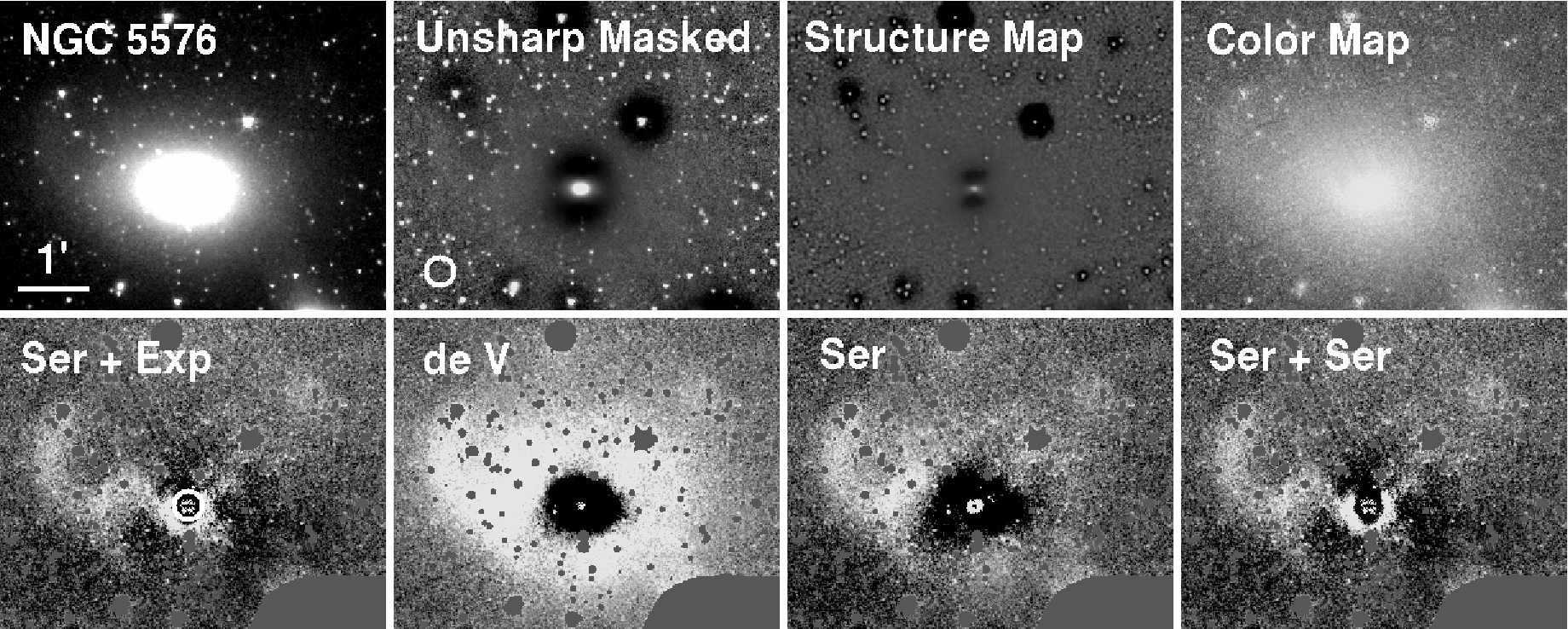}
\caption{Image analysis of NGC 5018 and NGC 5576, same as Figure \ref{ngc0474set6}.
\label{ngc5018set6} \label{ngc5576set6}}
\end{center}
\end{figure*}
%---------------------------------------------------------------
\begin{figure*}   
\begin{center} 
\includegraphics[width=12cm]{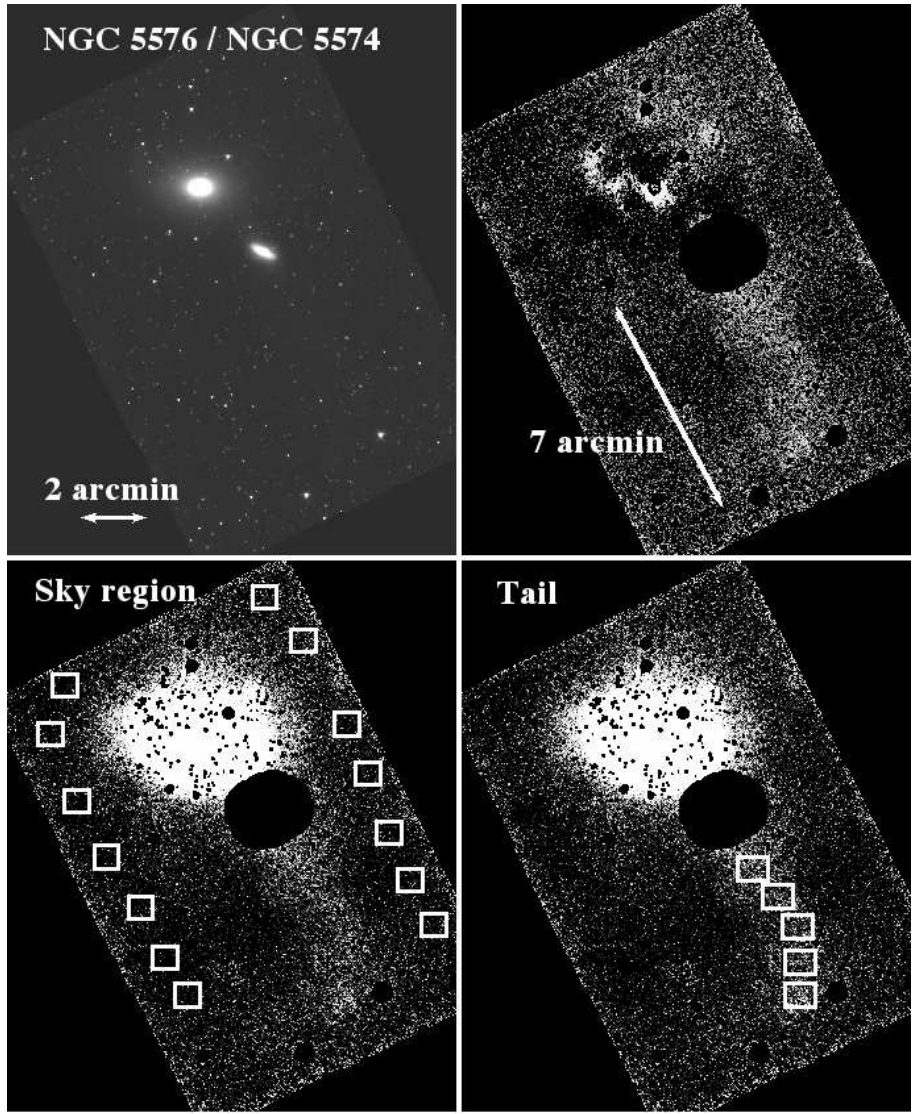}
\caption{The upper left panel shows the 3.6 {\micron} \s4g image of NGC 5576 (upper left galaxy) and NGC 5574 (lower right galaxy). The upper right panel shows the {\sc{GALFIT}} residual image of the model fit with {\ser}$+$exponential, displayed in histogram equalized stretch. A long tidal tail extends out to 7$\arcmin$ and tidal disturbances of NGC 5576 are also present.
The companion galaxy, NGC 5574, and stars were masked out to enhance faint structures.
The lower left panel presents the regions that were chosen to estimate the sky brightness value in the NGC 5576 field at 3.6 {\micron}, displayed in a histogram equalized stretch with foreground stars masked out. The lower right panel shows the regions that were adopted to compare the brightness of the tidal tail with that of the sky.
\label{ngc5576_big}}
\end{center}
\end{figure*}
%-------------------------------------------------------
\begin{figure*}
\begin{center}
\includegraphics[width=16cm]{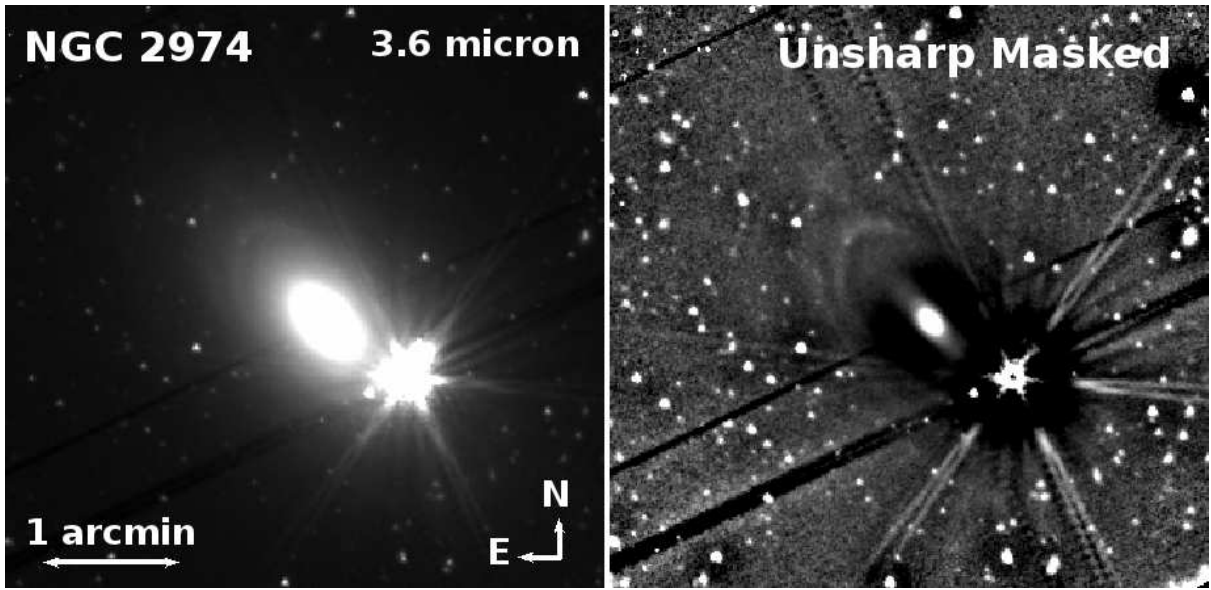}
\caption{
{\it Left}: NGC $2974$ at $3.6$ {\micron} {\it Right}: Unsharp masked image of NGC $2974$.} 
\label{ngc2974_usm}
\end{center}
\end{figure*}
%-------------------------------------------------------
\begin{figure*}
\begin{center}
\includegraphics[width=16cm]{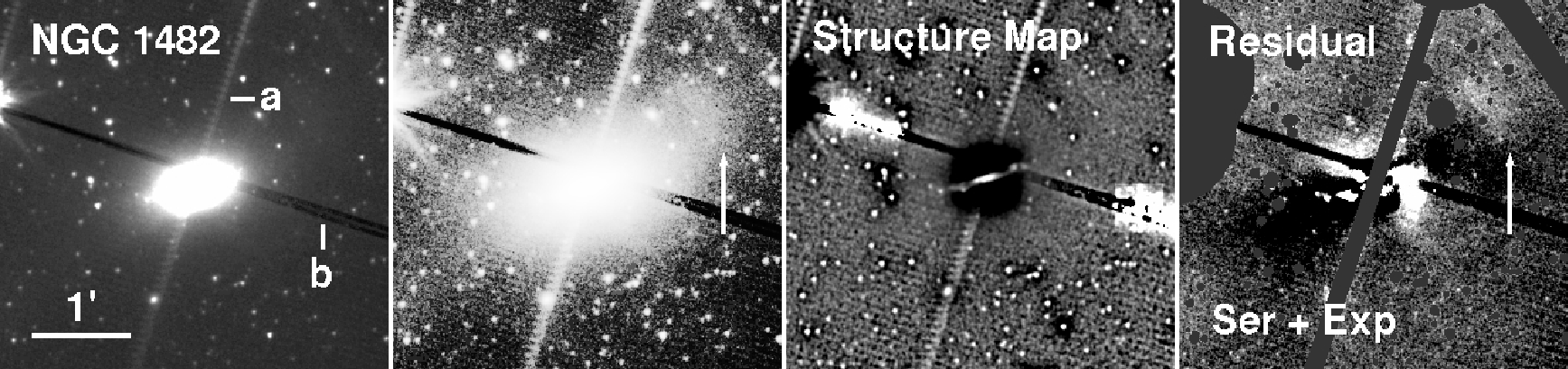}\\
\vspace{2 mm}
\includegraphics[width=16cm]{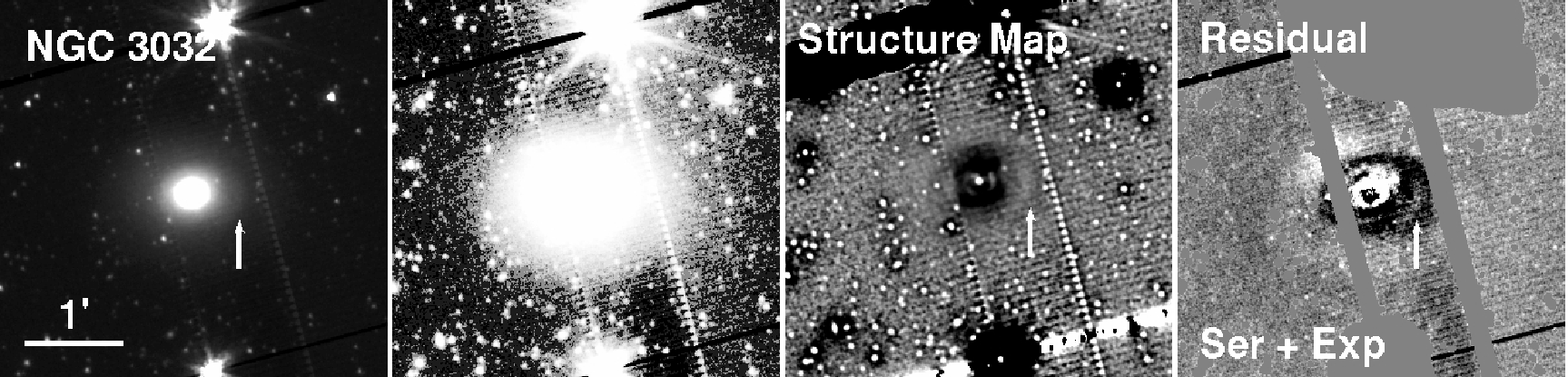} \\
\vspace{2 mm}
\includegraphics[width=16cm]{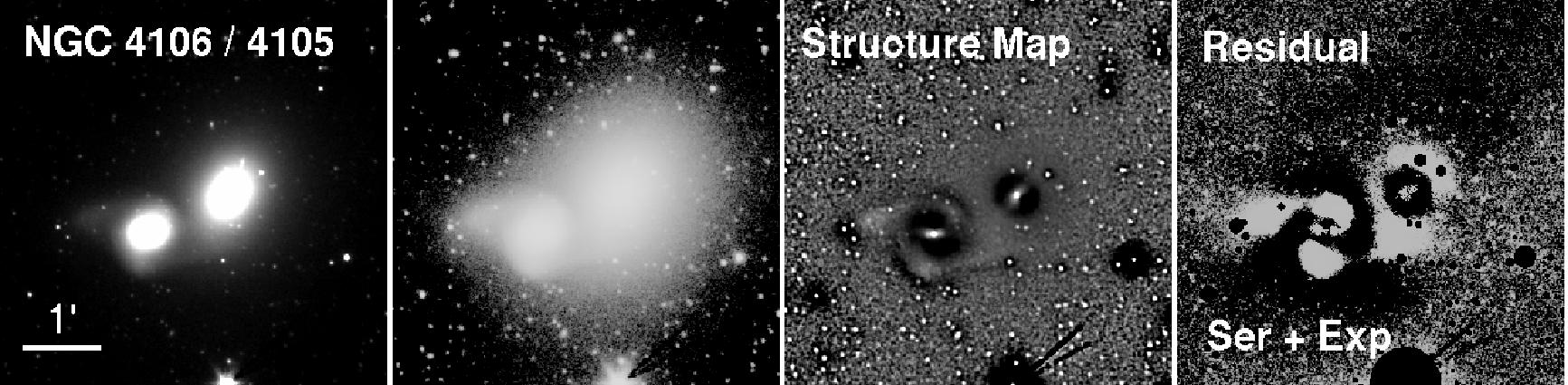}
\caption{Image analysis of NGC 1482, NGC 3032 and NGC 4106, which were identified to have tidal features. 
Note that these galaxies were not carried on for the further analysis, because of the muxbleed (marked with `a' in the upper panel) and column pull-down (marked with `b') due to saturation (NGC 1482 and NGC 3032) or close companion (NGC 4105 and NGC 4106). 
(\textit{Left to right}): $3.6$ {\micron} image in log-scale stretch; $3.6$ {\micron} image in histogram equalized stretch; structure map; {\sc{GALFIT}} residual images from a bulge (n=free) and an exponential disk (n=1) fit.
Tidal features are marked with arrows.
\label{ngc1482_set4}  \label{NGC3032_set4}   \label{NGC4106_set4} }
\end{center}
\end{figure*}

$\bullet$ NGC 1222: There is a tidal disturbance toward the North as can be seen in Figure \ref{ngc1222set6}. Also known as Markarian 604, NGC 1222 has a starburst nucleus, which has a compact radio and infrared source (\citealt{beck07}). \citet{petrosian93} claimed that this galaxy has appearance of a triple system in Figure 1 of their paper, and argued that two other galaxies are dwarf galaxies with high surface brightness interacting with the main galaxy. They also mentioned that there is a continuous twist in the isophotes. 
To check if we can confirm this isophotal twist, we have run the {\sc{ellipse}} task in {\sc{iraf}}, and we also find the twist in the isophotes of the galaxy.
Also NGC 1222 have a large kinematic misalignment between the photometric and kinematic major axes ($\Psi=72.7 \arcdeg$, where sin $\Psi = \mid$ sin(PA$_{phot} -$ PA$_{kin}) \mid$). Despite the complex morphology, NGC 1222 is classified as a slow rotator (\citealt{emsellem11}).

$\bullet$ NGC 1482: Toward the Northeast, the galaxy has a tidal feature, and it is marked with an arrow in Figure \ref{ngc1482_set4}. \citet{hameed99} note that this galaxy has ``filaments and/or chimneys of ionized gas extending perpendicular to the disk'' (see Figure 5 of their paper).  Large scale hourglass-shaped outflows has been detected with excitation maps of [N II] ${\lambda}6583$ and H${\alpha}$ also by \citet{veilleux02}. They claimed this galactic wind is extending along the minor axis of the galaxy at least 1.5 kpc above and below the galactic plane in both directions with a velocity of $\sim$250 \kms.

$\bullet$ NGC 4105/4106 pair: These galaxies are interacting and NGC 4106 (left one in Figure \ref{NGC4106_set4}) shows tidal disturbances. One of the tidal tails extends toward the elliptical galaxy, NGC 4105 (right one in Figure \ref{NGC4106_set4}). 
These galaxies are separated by a projected distance of 7.7 kpc and the difference in radial velocities is $\sim$260 \kms.
NGC 4106 shows a disky central structure  and NGC 4105 has dusty features in the central region and disky outer structure (\citealt{grutzbauch07}).
\citet{reduzzi96} obtained a nearly constant color profile of ($B-V$) $\sim$ 0.9 for both NGC 4105 and NGC 4106, while \citet{longhetti99} claimed that these galaxies experienced recent star formation episodes from their study of line indices. While NGC 4106 is perturbed, NGC 4105 does not show clear perturbation (\citealt{grutzbauch07}).
Thus, we exclude the elliptical galaxy, NGC 4105, from our selection of galaxies with tidal disturbances.

$\bullet$ NGC 4694: This galaxy is described as ``amorphous'' (\citealt{sandage81}), and
classified as an SB0(pec) in RC3 (\citealt{devaucouleurs91}). However, as can be seen in Figure \ref{ngc4694set6}, it is not easy to tell whether NGC 4694 exhibits any bar feature in the 3.6 {\micron} image. Using multi component galaxy decomposition, \citet{laurikainen10} note that it has a bright oval-like component and a peculiar nuclear bar or an inner disc. \citet{duc07} presented a VLA HI map of NGC 4694 and its nearby dwarf galaxy, VCC 2062, in their Figure 1. They show that the two galaxies are connected by an HI bridge. The faint spiral arm-like tidal feature of NGC 4694 in the bottom panel of Figure \ref{ngc4694set6} likely follows HI-arm. Therefore, this tidal feature may originate from an interaction with VCC 2062. 

$\bullet$ NGC 5576: NGC 5576 is reported to have boxy isophotes (\citealt{ebneter88}), and it shows peculiar tidal features, tidal disturbances to the North and a very long tidal stream to the South. The tidal disturbance in this galaxy can be seen in the right panel of Figure \ref{ngc5576_big}. At an angular separation of about 2.76$\arcmin$ ($\sim$ 16 kpc) to the South-west from NGC 5576 (V$_{rad}=1482$ \kms) there is a barred S0 type galaxy, NGC 5574 (V$_{rad}=1657$ \kms). There is a long tidal tail extending out to 60 kpc, which suggests that NGC 5576 and NGC 5574 are an interacting pair. To check how bright this tidal tail is compared to the sky, we chose square regions outside the galaxy, as shown in the bottom left panel of Figure \ref{ngc5576_big}. We then estimated the surface brightness of the tidal tail in these selected boxes.  The signal to noise ratio of the tidal tail region is low (S/N $\sim$1.5), which indicates that the tidal tail is only marginally brighter than the sky. \cite{tal09} also see this  faint tidal tail in their deep V-band image out to 75 kpc. 

Despite the faintness of the tidal tail, NGC 5576 and NGC 5574 are likely interacting. The nucleus of NGC 5574 has an iron-overabundant stellar system (\citealt{silchenko02}), which is a rare phenomenon for ellipticals, because most ellipticals are magnesium-overabundant. \cite{silchenko02} mention that this iron overabundance can be observed in some irregular galaxies and is usually interpreted as a result of multiple starbursts. Interestingly, the nucleus of the companion galaxy, NGC 5576, also shows this rare iron overabundance. The luminosity-weighted stellar ages of both galaxies are 3 Gyr (\citealt{silchenko97}), which implies that these two galaxies have experienced similar recent episodes of star formation. NGC 5576 is classified as a slow rotator (\citealt{emsellem11}) without any specific kinematic feature, such as kinematically distinct core or kinematic twist in the central part of the galaxy. This result implies that this galaxy experienced a significant merging event without being able to rebuild a fast rotating stellar component through dissipation (\citealt{emsellem11}). Although the long tidal tail seems to be related to the interaction of NGC 5576 and NGC 5574, and even if NGC 5576 shows tidal disturbances in the 3.6 {\micron} imaging data alone, we can not be sure from which galaxy the tidal tail originated. 
As catalogued in the Isolated triplets of galaxies (\citealt{karachentseva79}, KTG 55), NGC 5576 has another companion galaxy, NGC 5577, located to the North, which has a redshift similar to NGC 5576. Since we only cover the southern part of this galaxy pair at 3.6$\mu$m, we checked the northern part in the accompanying 4.5 {\micron} data, but could not identify a tidal bridge between NGC 5576 \& NGC 5577. Therefore, the tidal disturbance of NGC 5576 might be related to the interaction with NGC 5574, although NGC 5574 has a larger recession velocity or it may be that NGC 5574 has a high peculiar velocity due to its interaction with NGC 5576. 

$\bullet$ NGC 2974: This object is classified as an elliptical galaxy, E4 (\citealt{devaucouleurs91}), isolated elliptical (\citealt{sandage94}), and an SA$($r$)0/$a galaxy by \citet{buta10} using the \s4g data. \citet{tal09} identified multiple shells in this system. NGC 2974, however, shows an HI velocity field consistent with a rotating disk and the rotation axis of the galaxy is well aligned with the optical minor axis of the galaxy (\citealt{kim88}). Additionally, according to \citet{emsellem03}, this galaxy has gaseous spiral structure and a possible inner bar within a radius of $200$ pc.  From the near- and far-ultraviolet study (\citealt{jeong07}), NGC 2974 hosts an outer ring at a radius of 60$\arcsec$ with a young stellar population ($\le 1$ Gyr), and we confirm the arc-like ring features at a radius of 55$\arcsec$ (corresponds to 7 kpc) in the unsharp masked image of 3.6 {\micron} (Figure \ref{ngc2974_usm}). A bright star near NGC 2974 makes it difficult to determine whether this feature is a ring or a shell,  but the arc-like feature encompasses almost 180$\arcdeg$, which can be seen in Figure 1 of \citet{jeong07}. It also hosts a regular HI ring aligned with the stellar component of galaxy (\citealt{serra10}). Therefore, this arc-like feature may not be a shell as previously claimed (\citealt{tal09}). For these reasons, we excluded NGC 2974 from our sample of ETGs with tidal features.

\subsection{Fraction of tidally disturbed ETGs}
%::::::::::::::::::::::::::::::::::::::::::::::::::::::::::::::::::::::::::::::::::::::::::::::::::::
\setlength{\tabcolsep}{8pt}
\begin{deluxetable*}{lcccll}
\tablecolumns{6}
\tabletypesize{\scriptsize}
\tablewidth{0pc}
\tablecaption{Fraction of ETGs showing tidal features. \label{tab_fraction}}
\tablehead{ 
  \colhead{Study}    &
  \colhead{Fraction}&
  \colhead{Object}  &
  \colhead{Band}    &
  \colhead{Depth }  &
  \colhead{Equivalent depth in B} \\
   & \colhead{[$\%$]}  &  &  & \colhead{[mag]} & \colhead{[AB mag]}  \\
    \colhead{(1)} & \colhead{(2)} & \colhead{(3)} & \colhead{(4)} & \colhead{(5)} & \colhead{(6)}   } 
\startdata
\citet{malin83}           &   17\tablenotemark{a} (shell)  &  Ellipticals  &   B                      &  26.5 (Vega)                                         &   26.3                 \\
\citet{vandokkum05}  &   70              &  ETGs   &   BVR\tablenotemark{b}     &  29.5\tablenotemark{b} (AB)     $\to$ 28.0\tablenotemark{c}     &   29.5\tablenotemark{b}    $\to$  28.0\tablenotemark{c}      \\
                                  &                  &                &   BVI \tablenotemark{b}       &  29.0\tablenotemark{b} (AB)     $\to$ 28.0\tablenotemark{c}  &   29.0\tablenotemark{b}    $\to$ 28.0\tablenotemark{c}   \\
\citet{tal09}                &   73   &  ETGs       &   V                      &  29.0 (Vega)  $\to$ 27.7\tablenotemark{c} &    30.0   $\to$ 28.7\tablenotemark{c}    \\
\citet{krajnovic11}     &     8   &  ETGs        &   r                       &   26.0 (AB)                                           &   27.3  \\
 This study                 &  17$\pm 3$   &  ETGs        &   3.6${\:\mu}$m    &  27.0 (AB) $\to$ (25.2)\tablenotemark{e}            &  28.7   $\to$ (26.9)\tablenotemark{e}    \\
  &  13\tablenotemark{a}$\pm 1$ (shell) & Ellipticals &  & & \\
  &   6\tablenotemark{f}$\pm 2$ (shell) & ETGs & & &
 \enddata
\tablecomments{ 
Col. (1): Author of the study.
Col. (2): Fraction of tidally disturbed galaxies.
Col. (3): Morphology of parent sample.
Col. (4): Filter.
Col. (5): Typical 1 $\sigma$ surface brightness limit presented in the literature.
Col. (6): Typical 1 $\sigma$ surface brightness limit converted to B-band AB magnitude. We made use of the Table 13 of \citet{eisenhardt07} to convert magnitudes, and we took colors $(B-V)=1.0$, $(B-R)=1.5$ at H=13 mag. To convert r to R, we used the equation listed in SDSS web page.\tablenotemark{d} We took the value $(g-r)=0.7$ for ETGs and we used $(R-r)=-0.2$ for conversion. We adopt (3.6 {\micron} $-$ B) $\sim$1.7   as used in \citet{sheth10}.
}
\tablenotetext{a}{Fraction of shell galaxies in ellipticals galaxies.}
\tablenotetext{b}{Co-added.}
\tablenotetext{c}{Depth that \citet{vandokkum05} and \citet{tal09} can confidently identify features, mentioned in their paper.}
\tablenotetext{d}{$http://www.sdss.org/dr5/algorithms/sdssUBVRITransform.html\#Lupton2005$}
\tablenotetext{e}{The surface brightness of the tidal features in the shallowest image of our sample. With deeper images from the {\em{Spitzer}} Heritage Archive, we could identify tidal features down to $\mu_{3.6,AB} = 26.5$ ($\sim \: \mu_{B,AB} = 28.2$ ), see the text for detail.}
\tablenotetext{f}{Fraction of shell galaxies in ETGs, including elliptical and S0 galaxies.}
\end{deluxetable*}
%:::::::::::::::::::::::::::::::::::::::::::::::::::::::::::::::::::::::::::::::::::::::::::::::::::::

We have identified 11 tidally disturbed galaxies from 65 ETGs. Note that our sample of this pilot study does not include the whole statistically complete sample of ETGs in \s4g. We found fewer ETGs with tidal features than \citet{vandokkum05} and \citet{tal09}, who found fine structures in 70$\%$ of their ETGs.
Tidal features can be relatively faint, as shown by \citet{janowiecki10},  the peak surface brightness of tidal features range from $\mu_{V}$ = 26 -- 29 mag arcsec$^{-2}$.
Therefore, the ability to identify tidal features depends strongly on the data, especially on the depth of the image.
We summarize the fraction of tidally disturbed galaxies from the literature in Table \ref{tab_fraction} along with the depth of images analyzed in each study. 
As studies in the literature were performed in different bands, we convert the magnitude of depth of images to B-band AB magnitude system in Table \ref{tab_fraction} to directly compare the fraction of tidally disturbed ETGs. As we can see in Table \ref{tab_fraction}, the deeper studies can find more tidally disturbed ETGs.
Depending on the project, the {\em{Spitzer}} archival images have different depths and we could find tidal features down to $\mu_{3.6,AB}\leq26.5$ mag arcsec$^{-2}$, while in the newly obtained ``Warm'' data from the \s4g project, we found tidal features down to $\mu_{3.6,AB}\leq25.2$ mag arcsec$^{-2}$ ($\sim\mu_{B,AB}{\leq}$ 26.9 mag arcsec$^{-2}$). 

\s4g maps each galaxy to 1.5$\times D_{25}$. However, for one of our sample galaxies, NGC 5846, which is identified to have shells by \citet{tal09}, we could not identify those shells because they lie outside of our field. As discussed in Section 3.2, we exclude NGC 2974 when we count ETGs showing shells or tidal disturbances.
IC 51 (\citealt{malin83}) and NGC 3608 (\citealt{forbes92}) are found to have shells from the literature\footnote{NGC 4494 is in the sample of shell galaxies in Table 4 of \citet{forbes94}, but afterwards there was no more description on the shell in this galaxy (\citealt{forbes95b}, \citealt{foster11}). So we exclude NGC 4494 from the sample that have been reported to have a shell in the optical.}.  But, we did not find clear evidence for tidal features.  With magnitude and field of view limit, we find 17${\pm}$3\% (11/65) of ETGs are tidally disturbed and 6${\pm}$2\% (4/65) have shell. If we restrict our sample to ellipticals using the morphology information of RC3, we find that 13${\pm}$1\% (2/19) of ellipticals have shells to our limiting surface brightness and field of view. We make use of Poisson statistics to present an uncertainty estimate of our fraction, using
 $\sigma=[f(1-f/N)]^{1/2}$ where N is the sample size and f is the number of galaxies showing tidal debris or shells following the method of \citet{laine02}.

\section{Luminosity of tidal features}
To estimate the luminosity of the tidal features, we first mask out all the foreground stars and background galaxies, and after checking the residual image of each galaxy, we also mask out known features such as lens components or an inner bar, mentioned in literature. The masked out inner and outer boundaries are listed in Table \ref{tab_ratio_result}.
Because now we focus on the tidal features, this time we do not mask out tidal features but leave them in the {\sc{GALFIT}} residual image.
Then, we estimate the ratio of the total counts from the tidal debris in the residual image to the total counts in the model image.
 We calculated the luminosity fraction of the tidal features as a percentage, 
 ($F_{residual}/F_{model}$) $\times$ 100 and list these in Table \ref{tab_ratio_result}.
 The luminosity fractions of tidal features are estimated both in a circular region and in an elliptical region that is chosen to have the same PA and ellipticity as the bulge component of the model has. The ratios using the circular regions are labelled as $Ratio_{circ}$, and those using the elliptical region are labelled as $Ratio_{ell}$.  
In the cases of the {\ser}$+$exponential fit and the {\ser}$+${\ser} fit, the fractional luminosity is estimated using the total model, which is the sum of both the bulge and the disk. For simplicity, we present these ratios in the first row of S$+$E and S$+$S in Table \ref{tab_ratio_result}.

Some of our sample galaxies (NGC 1482 and NGC 3032) are severely saturated and have the ``muxbleed'' and ``column pull-down'' artifacts which result from a bright source -- see the IRAC Instrumental Handbook Section 7.2 for more details. The ``muxbleed'' is a trail of bright pixels along a row in the readout direction that repeats with
an increased signal every fourth column. This is denoted as `a' in NGC 1482 of Figure \ref{ngc1482_set4}. The ``column pull-down'' artifact has a decreased signal above and below the bright source along a column. This is denoted as `b' in Figure \ref{ngc1482_set4}. Note that we excluded galaxies which were severely contaminated by these effects in our analysis. Close companions (NGC 4105/ 4106 pair) that lie within 1/4 of $D_{25}$ were also excluded. 
 
If we include all the pixel values in the residual images to estimate total counts of tidal features, then we measure the luminosity fraction of the tidal feature to be 0.5 -- 7\% of the total model galaxy luminosity. But as described in \citet{bennert08}, to avoid adding negative values from the regions that were oversubtracted by {\sc{GALFIT}} in the residual image, now we only sum values of all pixels brighter than $-1\sigma$ of the sky background to get total counts, and list luminosity fractions in Table \ref{tab_ratio_result}. 
If we only sum values of pixels that are brighter than $-3\sigma$ of the sky background, then we obtain the luminosity of 1 -- 8\% of the total galaxy luminosity.
For galaxies with shells, this result is consistent with those of \citet{prieur88}, \citet{canalizo07}, and \citet{bennert08}, who found that the shells constitute 5\% of B-band, $\sim$ 6\% of F606W-band, and 5 -- 10\% of F606W-band luminosity of the galaxy, respectively. We also estimate the tidal parameter, which were used in \citet{vandokkum05} and \citet{tal09}. The tidal parameter is defined as,
\begin{equation}
 T_{galaxy}=\overline{\vert I_{x,y}-M_{x,y} \vert},
\end{equation}
where $I_{x,y}$ and $M_{x,y}$ are pixel values at the position x, y of the object and model, respectively. The tidal parameters of our sample galaxies range from 0.09 to 0.23. These values are comparable to what \citet{vandokkum05} and \citet{tal09} found for galaxies that have tidal features.

In Section 3.2, for the galaxies which show isophotal twist or asymmetric envelope, 
we have run the {\sc{Ellipse}} task in {\sc{iraf}} to confirm the presence of those features. We also estimate the luminosity of tidal features using the fit from {\sc{Ellipse}}. The luminosity of the tidal features does not change significantly in the case of NGC 680, but is higher in the case of NGC 1222 with ellipse fits. Because of the artificial features shown in the models derived from {\sc{Ellipse}} fits, and for the sake of consistency, we will only consider here the results derived from the {\sc{GALFIT}} models.

By assuming the same stellar $M/L$ in shells and galaxies, we estimate that 3\% to 10\% of the galaxy stellar mass is contained in the shells or tidal features.
The amount of mass in shells can be used to infer the type of merger that led to its creation. \citet{canalizo07} discuss that with their own numerical simulations and those of \citet{hernquist92}, if shell features were formed through major mergers, the stars in the resulting shells constitute one fourth or less of the total mass of the companion galaxy. Following the argument of \cite{canalizo07}, shells with 5\% of the mass are likely to result from a merger with a 3:1 mass ratio, whereas shells with 10\% of the mass are more likely to come from a merger where the galaxies have comparable masses. 

The luminosity of the long tidal stream of the NGC 5576/5574 pair is estimated in the same manner. We find that the long tidal stream constitutes 4$\%$ of the galaxy model luminosity of NGC 5576. The estimated luminosity of the tidal features here might change slightly due to foreground stars or background galaxies that might not be fully masked. We increase the size of mask by 3 pixels for every foreground stars and background galaxies, and we find that the amount that long tidal stream constitutes decreased to 3$\%$ of the galaxy model luminosity of NGC 5576.

%-------------------------------------------------------
\setlength{\tabcolsep}{1.9pt}
\begin{deluxetable*}{llcccccccccccccc}
\tablecolumns{16}
\tabletypesize{\scriptsize}
\tablewidth{0pc}
\tablecaption{Results of modeling galaxies\label{tab_ratio_result}}
\tablehead{  
 \colhead{Object} &
 \colhead{Model} &
 \colhead{n}&
 \colhead{type}&
 \colhead{$r_{\rm eff}$}&
 \colhead{$r_{\rm eff}$}&
 \colhead{$h_{\rm r}$}&
 \colhead{$h_{\rm r}$}&
 \colhead{PA}&
 \colhead{b/a}&
 \colhead{$mag_{3.6,AB}$}&
 \colhead{$Mag_{3.6,AB}$} &
 \colhead{$Ratio_{circ}$} &
 \colhead{$Ratio_{ell}$} &
 \colhead{B/T}&
 \colhead{${{\chi}_{\nu}}^2$}  \\
 \colhead{[inner,outer]} & & & & \colhead{[$\arcsec$]}& \colhead{[kpc]} & \colhead{[$\arcsec$]}& \colhead{[kpc]} &
\colhead{[$\arcdeg$]} & & \colhead{[mag]} & \colhead{[mag]} & (Circular) & (Elliptical)& & \\
\colhead{(1)} & \colhead{(2)} & \colhead{(3)} & \colhead{(4)} & \colhead{(5)} & \colhead{(6)} & \colhead{(7)} & \colhead{(8)} & \colhead{(9)} & \colhead{(10)} & \colhead{(11)} & \colhead{(12)} & \colhead{(13)} & \colhead{(14)} & \colhead{(15)} & \colhead{(16)}}
\startdata
%-------------------------------------------------------------------------------------------
         NGC 0474 &                   S   & 6.65 &     free &   31.78 &      4.86 &   \nodata &   \nodata &    11.3 &      0.91 &    10.88 &   -21.65 &      6.2 &      6.3 &  \nodata   &     1.14\\
   $[45.0,217.5]$ &              S+E (B)  & 2.44 &     free &    4.46 &      0.68 &   \nodata &   \nodata &     8.3 &      0.90 &    12.14 &   -20.39 &      11.0 &      11.1  &     0.39   &     0.82\\
          \nodata &              S+E (D)  & 1.00 &      fix &   40.30 &      6.16 &     24.02 &      3.67 &   -15.2 &      0.94 &    11.67 &   -20.85 &  \nodata &  \nodata &  \nodata   &  \nodata\\
          \nodata &                  DeV  & 4.00 &      fix &   14.29 &      2.19 &   \nodata &   \nodata &    12.7 &      0.93 &    11.23 &   -21.30 &     19.1  &     18.5 &  \nodata   &     1.90\\
          \nodata &              S+S (B)  & 2.34 &     free &    4.18 &      0.64 &   \nodata &   \nodata &     7.7 &      0.90 &    12.20 &   -20.33 &      10.4 &      10.5 &     0.37   &     0.81\\
          \nodata &              S+S (D)  & 1.12 &     free &   39.81 &      6.09 &   \nodata  &  \nodata  &   -10.6 &      0.94 &    11.62 &   -20.91 &  \nodata &  \nodata &  \nodata   &  \nodata\\
\tableline
         NGC 0680 &                   S   & 5.76 &     free &   15.93 &      2.93 &   \nodata &   \nodata &   -10.5 &      0.78 &    11.06 &   -21.87 &      5.5 &      5.4 &  \nodata   &     2.44\\
   $[12.8,127.5]$ &              S+E (B)  & 3.38 &     free &    4.91 &      0.90 &   \nodata &   \nodata &    -4.3 &      0.78 &    11.84 &   -21.09 &      6.7 &      7.2 &     0.55   &     2.06\\
          \nodata &              S+E (D)  & 1.00 &      fix &   32.17 &      5.91 &     19.17 &      3.52 &   -23.7 &      0.75 &    12.07 &   -20.86 &  \nodata &  \nodata &  \nodata   &  \nodata\\
          \nodata &                  DeV  & 4.00 &      fix &   10.55 &      1.94 &   \nodata &   \nodata &    -9.0 &      0.78 &    11.23 &   -21.71 &     12.2 &     12.5 &  \nodata   &     3.75\\
          \nodata &              S+S (B)  & 2.57 &     free &    2.98 &      0.55 &   \nodata &   \nodata &     0.2 &      0.77 &    12.28 &   -20.65 &      5.7 &      6.1 &     0.36   &     2.01\\
          \nodata &              S+S (D)  & 1.63 &     free &   26.47 &      4.87 &   \nodata &   \nodata &   -21.0 &      0.76 &    11.66 &   -21.27 &  \nodata &  \nodata &  \nodata   &  \nodata\\
\tableline
         NGC 1222 &                   S   & 3.25 &     free &    5.82 &      0.94 &   \nodata &   \nodata &   -73.5 &      0.68 &    12.05 &   -20.59 &     14.6 &     15.5 &  \nodata   &     3.84\\
   $[24.6,120.0]$ &              S+E (B)  & 2.01 &     free &    3.73 &      0.60 &   \nodata &   \nodata &   -78.1 &      0.59 &    12.42 &   -20.22 &      7.3 &      8.8 &     0.66   &     2.58\\
          \nodata &              S+E (D)  & 1.00 &      fix &   28.01 &      4.51 &     16.69 &      2.69 &   -15.6 &      0.61 &    13.16 &   -19.48 &  \nodata &  \nodata &  \nodata   &  \nodata\\
          \nodata &                  DeV  & 4.00 &      fix &    6.70 &      1.08 &   \nodata &   \nodata &   -72.4 &      0.69 &    11.99 &   -20.65 &     11.5 &     12.5 &  \nodata   &     4.06\\
          \nodata &              S+S (B)  & 2.82 &     free &   21.37 &      3.44 &   \nodata &   \nodata &   -19.4 &      0.68 &    12.63 &   -20.01 &      5.1 &      6.2 &     0.51   &     2.41\\
          \nodata &              S+S (D)  & 1.89 &     free &    3.43 &      0.55 &    \nodata &   \nodata &   -80.6 &      0.50 &    12.69 &   -19.95 &  \nodata &  \nodata &  \nodata   &  \nodata\\
\tableline
         NGC 2634 &                   S   & 6.41 &     free &   29.59 &      4.40 &   \nodata &   \nodata &    32.7 &      0.91 &    11.22 &   -21.25 &      3.9  &      4.3 &  \nodata   &     1.23\\
   $[12.8,142.5]$ &              S+E (B)  & 4.06 &     free &    9.25 &      1.38 &   \nodata &   \nodata &    29.1 &      0.88 &    11.96 &   -20.50 &   5.0 &      6.1 &     0.61   &     1.17\\
          \nodata &              S+E (D)  & 1.00 &      fix &   47.61 &      7.08 &     28.38 &      4.22 &    74.7 &      0.95 &    12.47 &   -20.00 &  \nodata &  \nodata &  \nodata   &  \nodata\\
          \nodata &                  DeV  & 4.00 &      fix &   15.23 &      2.26 &   \nodata &   \nodata &    33.1 &      0.91 &    11.49 &   -20.97 &     13.7 &     14.5 &  \nodata   &     1.92\\
          \nodata &              S+S (B)\tablenotemark{a}  & 2.97 &     free &   32.41 &      4.82 &   \nodata &   \nodata &    34.5 &      0.92 &    11.55 &   -20.91 &      3.9 &      4.4 &     0.83   &     1.12\\
          \nodata &              S+S (D)\tablenotemark{a}  & 1.90 &     free &    2.38 &      0.35 &  \nodata  &    \nodata  &    29.4 &      0.88 &    13.31 &   -19.16 &  \nodata &  \nodata &  \nodata   &  \nodata\\
\tableline
         NGC 4694 &                   S   & 3.53 &     free &   32.92 &      2.52 &   \nodata &   \nodata &   -34.9 &      0.46 &    11.23 &   -19.78 &     2.6  &   3.6 &  \nodata   &     1.20\\
   $[69.8,187.5]$ &              S+E (B)  & 3.23 &     free &   25.84 &      1.98 &   \nodata &   \nodata &   -34.3 &      0.42 &    11.49 &   -19.52 &   2.8 &   4.3 &     0.82   &     1.06\\
          \nodata &              S+E (D)  & 1.00 &      fix &   51.23 &      3.93 &     30.53 &      2.34 &   -62.5 &      0.79 &    13.12 &   -17.89 &  \nodata &  \nodata &  \nodata   &  \nodata\\
          \nodata &                  DeV  & 4.00 &      fix &   37.93 &      2.91 &   \nodata &   \nodata &   -34.9 &      0.46 &    11.17 &   -19.84 &     1.7 &    1.6 &  \nodata   &     1.29\\
          \nodata &              S+S (B)  & 3.25 &     free &   27.12 &      2.08 &   \nodata &   \nodata &   -34.5 &      0.43 &    11.42 &   -19.59 &   3.1 &    4.4  &     0.87   &     1.05\\
          \nodata &              S+S (D)  & 0.61 &     free &   52.81 &      4.05 &  \nodata  &   \nodata  &   -82.6 &      0.79 &    13.51 &   -17.50 &  \nodata &  \nodata &  \nodata   &  \nodata\\
\tableline
         NGC 5018 &                   S   & 5.39 &     free &   23.83 &      4.41 &   \nodata &   \nodata &   -83.4 &      0.71 &     9.99 &   -22.96 &     2.6  &  2.6  &  \nodata   &     1.26\\
   $[19.6,247.5]$ &              S+E (B)  & 5.29 &     free &   19.08 &      3.53 &   \nodata &   \nodata &   -84.4 &      0.67 &    10.22 &   -22.73 &   4.3 &     4.1 &     0.87   &     1.08\\
          \nodata &              S+E (D)  & 1.00 &      fix &   30.73 &      5.69 &     18.31 &      3.39 &   -30.8 &      0.90 &    12.26 &   -20.68 &  \nodata &  \nodata &  \nodata   &  \nodata\\
          \nodata &                  DeV  & 4.00 &      fix &   16.93 &      3.13 &   \nodata &   \nodata &   -84.1 &      0.72 &    10.12 &   -22.83 &    8.9 &    8.2 &  \nodata   &     2.20\\
          \nodata &              S+S (B)  & 5.13 &     free &   11.23 &      2.08 &   \nodata &   \nodata &   -86.6 &      0.64 &    10.82 &   -22.13 &    4.7 &   4.5 &     0.51   &     1.06\\
          \nodata &              S+S (D)  & 2.45 &     free &   31.55 &      5.83 &   \nodata  & \nodata  &   -75.8 &      0.78 &    10.86 &   -22.09 &  \nodata &  \nodata &  \nodata   &  \nodata\\
\tableline
         NGC 5576 &                   S   & 6.11 &     free &   31.07 &      3.05 &   \nodata &   \nodata &    87.7 &      0.70 &    10.10 &   -21.45 &   4.9 &    4.9 &  \nodata   &     1.48\\
   $[11.8,247.5]$ &           S+E (B)  & 3.93 &     free &   11.82 &      1.16 &   \nodata &   \nodata &    89.2 &      0.69 &    10.68 &   -20.88 &  4.7 &  5.1  &     0.65   &     1.05\\
          \nodata &              S+E (D)  & 1.00 &      fix &   78.84 &      7.74 &     46.98 &      4.61 &    79.9 &      0.72 &    11.33 &   -20.22 &  \nodata &  \nodata &  \nodata   &  \nodata\\
          \nodata &                  DeV  & 4.00 &      fix &   17.44 &      1.71 &   \nodata &   \nodata &    87.9 &      0.72 &    10.33 &   -21.23 &     17.1 &     17.0 &  \nodata   &     3.61\\
          \nodata &              S+S (B)  & 3.45 &     free &    8.89 &      0.87 &   \nodata &   \nodata &    89.6 &      0.69 &    10.90 &   -20.66 &     4.4 &      4.7 &     0.52   &     1.04\\
          \nodata &              S+S (D)  & 1.48 &     free &   71.10 &      6.98 &  \nodata  &  \nodata  &    82.2 &      0.72 &    10.98 &   -20.57 &  \nodata &  \nodata &  \nodata   &  \nodata 
\enddata

\tablecomments{
Col. (1): Object name followed by inner and outer boundaries in arcsecs within square brackets, used to estimate the luminosity of the tidal features.
Col. (2): The models used to fit galaxies with {\sc{GALFIT}}. S represents the {\ser}, E represents the exponential and DeV is the de Vaucouleurs profile. B and D in parentheses  denote the bulge and disk component, respectively.
Col. (3): {\ser} index.
Col. (4): Type of the fit while running {\sc{GALFIT}} for each component. The Exponential and de Vaucouleurs profile are fixed to have n=1 and 4, respectively and denoted as ``fix''.
Col. (5): Effective radius measured along the semi major axis in arcsecs for bulge and also for disk.
Note that for exponential disks, the scale length can be obtained by dividing this effective radius by 1.678.
Col. (6): Effective radius measured along the semi major axis in kpc, estimated with the heliocentric redshift listed in NED.
Col. (7): Disk scale length in arcsecs for disk component.
Col. (8): Disk scale length in kpc for disk component.
Col. (9): Position Angle, east of north. (North = 0, East = 90)
Col. (10): Minor axis to major axis ratio.
Col. (11): Apparent AB magnitude at 3.6 {\micron}.
Col. (12): Absolute AB magnitude at 3.6 {\micron}.
Col. (13): The ratio of the residual to the galaxy model in the circular region between inner and outer boundary. Note that in the case of {\ser} with exponential fit, we estimate the ratio with the total model of bulge and disk, but we put the value just in the first row in the S$+$E section.
Col. (14): The ratio of the residual to the galaxy model in an elliptical boundary using PA, b/a, inner and outer boundary.
Col. (15): Bulge to Total ratio.
Col. (16): The reduced ${{\chi}^2}$, which is a indicator of goodness of fit.
} 
\tablenotetext{a}{From the {\ser}+{\ser}, two-component fit, we took the larger {\ser} index component as a bulge component. But in this case (NGC 2634), the effective radius of the larger {\ser} index component (bulge) is larger than that of smaller {\ser} index component (disk), hence this model fit may not be realistic. }
\end{deluxetable*}
%-----------------------------------------------------------------------------

\section{Structural Properties}
\subsection{Modeling light distribution of galaxies with different functions}
The structural properties such as sersic index, effective radius, disk scale length, position angle, minor to major axis ratio, and bulge to total of ETGs having tidal features are presented in Table \ref{tab_ratio_result}.  As noted in Section 2.2, we perform {\sc{GALFIT}} twice, the preliminary {\sc{GALFIT}} run is on the image where foreground stars and background galaxies were masked out and the second {\sc{GALFIT}} run is on the image where tidal features were also masked out. The results in Table \ref{tab_ratio_result} are obtained with the images where tidal features were also masked out. The differences between two different runs show that the {\ser} indices, effective radii of bulges, and scale lengths of disks decrease by less than 10\%, except for NGC 474, where the effective radius of the bulge component that was obtained from the {\ser} only model fit decreases by ${\sim}$ 20\% from the first preliminary {\sc{GALFIT}} run to the second {\sc{GALFIT}} run.

\begin{figure*}
\begin{center}
\epsscale{0.85}
\includegraphics[width=16cm]{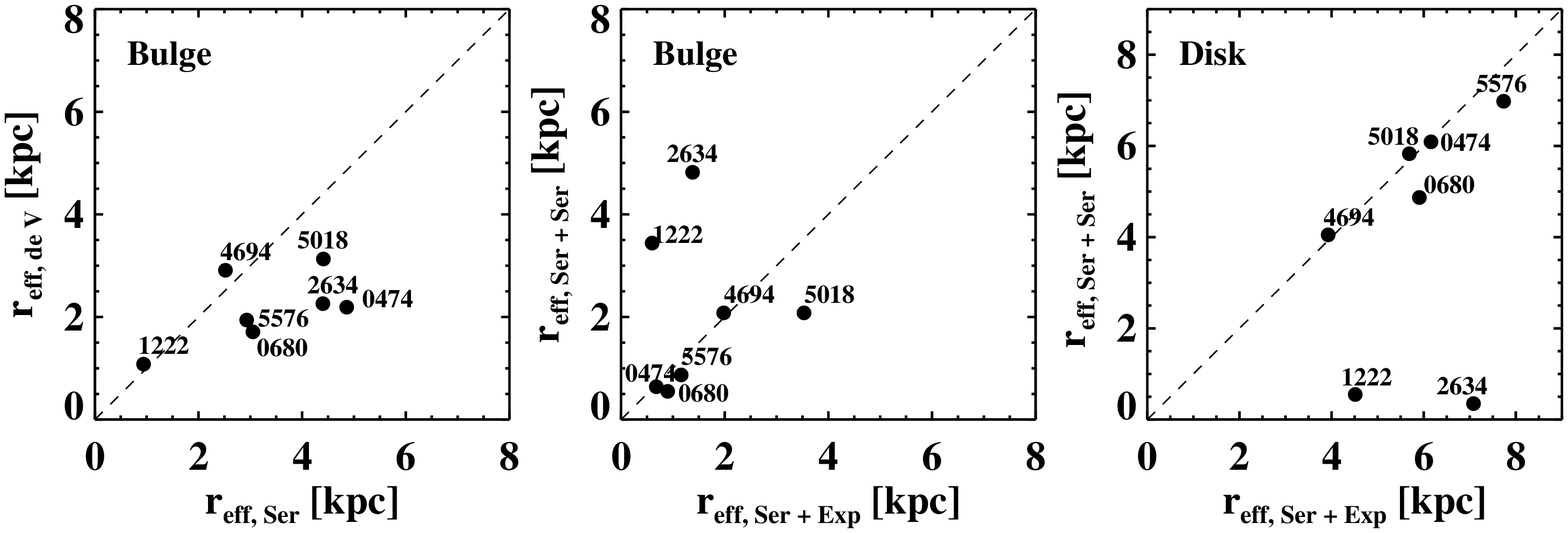}
\caption{Comparison of structural parameters among different choices of models. The numbers next to each point are the NGC numbers of the sample galaxies investigated in this study. The left panel shows differences in the effective radii of the bulge obtained from {\ser} (n=free) and de Vaucouleurs $R^{1/4}$ model fit. The middle panel compares the effective radii of the bulge estimated from the fit of {\ser}+exponential disk and {\ser}+{\ser} model fit.  The right panel compares the effective radii of the disk component obtained from the fit of {\ser}+exponential model fit and {\ser}+{\ser} model. Dashed lines guide one to one correspondence. 
Note  that the {\ser}+{\ser} model fit of NGC 1222 and NGC 2634 may not be realistic, as denoted in Table \ref{tab_ratio_result}.
}\label{compare_st}
\end{center}
\end{figure*}

We exclude galaxies that have muxbleeds and close companions from our final sample as discussed in the previous section and present the results for seven ETGs. Because not all ETGs follow the $R^{1/4}$ law, we need to test several model profiles to see which one best describes the two-dimensional light distribution of these galaxies.  
Even though galaxies were once classified as ellipticals, some of those ellipticals might actually be misclassified S0s.  By including a disk component in the model, we increase our chance of diagnosing such cases. Some of those examples are NGC 2634 and NGC 5018. They were first classified as ellipticals in the RC3 using the optical band, but recently they were reclassified as S0s (\citealt{buta10}) from 3.6 {\micron} data. In fact, except for NGC 680 and NGC 5576, all galaxies are classified as S0s, and therefore, we also present disk scale lengths of these galaxies in Table \ref{tab_ratio_result}. We see that parameters such as the {\ser} index, the effective radius or the disk scale length change with chosen model. We compare the differences in the effective radii in Figure \ref{compare_st}. The left panel shows differences in the effective radii obtained from the {\ser} model fit and the de Vaucouleurs $R^{1/4}$ model fit. For most galaxies, the effective radii obtained from the de Vaucouleurs $R^{1/4}$ model fit are smaller than those from the {\ser} model fit and they cluster between 2--3 kpc. The middle panel of Figure \ref{compare_st} compares the effective radii of the bulge components, estimated from a fit of {\ser} (for bulge, n=free) $+$ exponential disk model fit and {\ser} (for bulge, n=free)$+${\ser} (for disk, n=free) model. For the {\ser} (n=free)$+${\ser} (n=free) model, we adopt the component with a larger {\ser} index as a bulge and the one with smaller {\ser} index as a disk. For NGC 1222 and NGC 2634, the effective radius of the larger {\ser} index component (bulge) is {\em larger} than that of the smaller {\ser} index component (disk). Therefore, for those two galaxies, this model profile fit may not be realistic, but for reference we keep those data in Figure \ref{compare_st}, \ref{kormendy_rel}, and \ref{FP}. If we introduce a disk component to our model galaxies, the effective radius of the bulge component becomes smaller. We see this in the distribution of effective radius in the left and middle panel. Therefore, we must be careful not to blindly classify all galaxies as ETGs and treat them as a group, but we must model the lenticular galaxies including a disk component.  

If we let {\sc{GALFIT}} model a disk component with a free {\ser} index, we find that the {\ser} index of the disk is never equal to 1, as would be expected for a conventional exponential disk. This may be because disks of galaxies cannot be characterized by a single {\ser} index (\citealt{freeman70}).  In spiral galaxies, the outer disks show truncations (down-bending) or anti-truncations (up-bending) profiles (e.g., \citealt{pohlen02}, \citealt{erwin05}, \citealt{munoz-mateos09a}, Mu\~noz-Mateos et al. 2012, in preparation). Even though the {\ser} index deviates from n=1, the effective radius of the exponential disk from the {\ser}$+$exponential model fit and {\ser}$+${\ser} model fit is quite similar, as seen in the right panel of Figure \ref{compare_st}.
  
We tabulate kinematic class (\citealt{emsellem11}) and kinematic misalignment angle (\citealt{krajnovic11}) in Table \ref{tab_basicinfo}. If galaxies show tidal features, there should be imprints on the kinematics of those galaxies, and we check the result of our sample galaxies against those of the ATLAS$^{3D}$ study. 
NGC 474, NGC 680 and NGC 1222 show a rather large kinematic misalignment ($\Psi > 20 \arcdeg$) between the photometric and kinematic major axes. Only 10\% (25/260) of ATLAS$^{3D}$ galaxies have $\Psi > 20 \arcdeg$ and these are extreme cases. 
We find that one of our shell galaxies (NGC 474) and the galaxies with tidal disturbances (NGC 680 and NGC 4694) are better fitted with a disk component - these galaxies were classified as fast rotators (\citealt{emsellem11}). This result is consistent with the finding of \citet{cappellari11b} that fast rotators are generally lenticular galaxies. 
One exception is NGC 5576. We find that this galaxy is better described with disk component, but it is classified as a slow rotator (\citealt{emsellem11}) and according to \citet{krajnovic11}, it is classified as a non-regular rotator, which does not show any specific feature on the velocity map, such as kinematically distinct core or kinematic twist or counter rotating core. But \citet{krajnovic11} also note that it is possible that in lower noise velocity map, the galaxy would be classified as a regular rotator, which implies it might have a disk component. Thus our results from model fits are in agreement with the kinematic study from ATLAS$^{3D}$.
  
We find that the total magnitudes that were obtained assuming a specific model fit can vary among the different fits.
For example, the total magnitude of NGC 474 varies up to 0.3 magnitudes between n$=$4, de Vaucouleurs' fit and n=free {\ser} fit.
If a galaxy is better described by a n$<$4 fit, but modeled with n$=$4, then we obtain a brighter total magnitude. On the other hand, if a galaxy is better described by a n$>$4 fit, but modeled with n$=$4, then we obtain a fainter total magnitude.
For NGC 1222 and NGC 4694, {\ser} indices of one-component {\ser} fits are smaller than 4, and the total magnitudes obtained with those {\ser} models are larger than those from de Vaucouleurs' fits. For other galaxies that have n$>$4, the opposite holds. 
Studies of ETGs often estimate total magnitudes assuming their light profiles as de Vaucouleurs' law. But, as shown here, it is important to consider the differences due to the fitting model to obtain reliable estimates.

\subsection{Scaling relations}
\begin{figure*}
\begin{center}
\includegraphics[width=14cm]{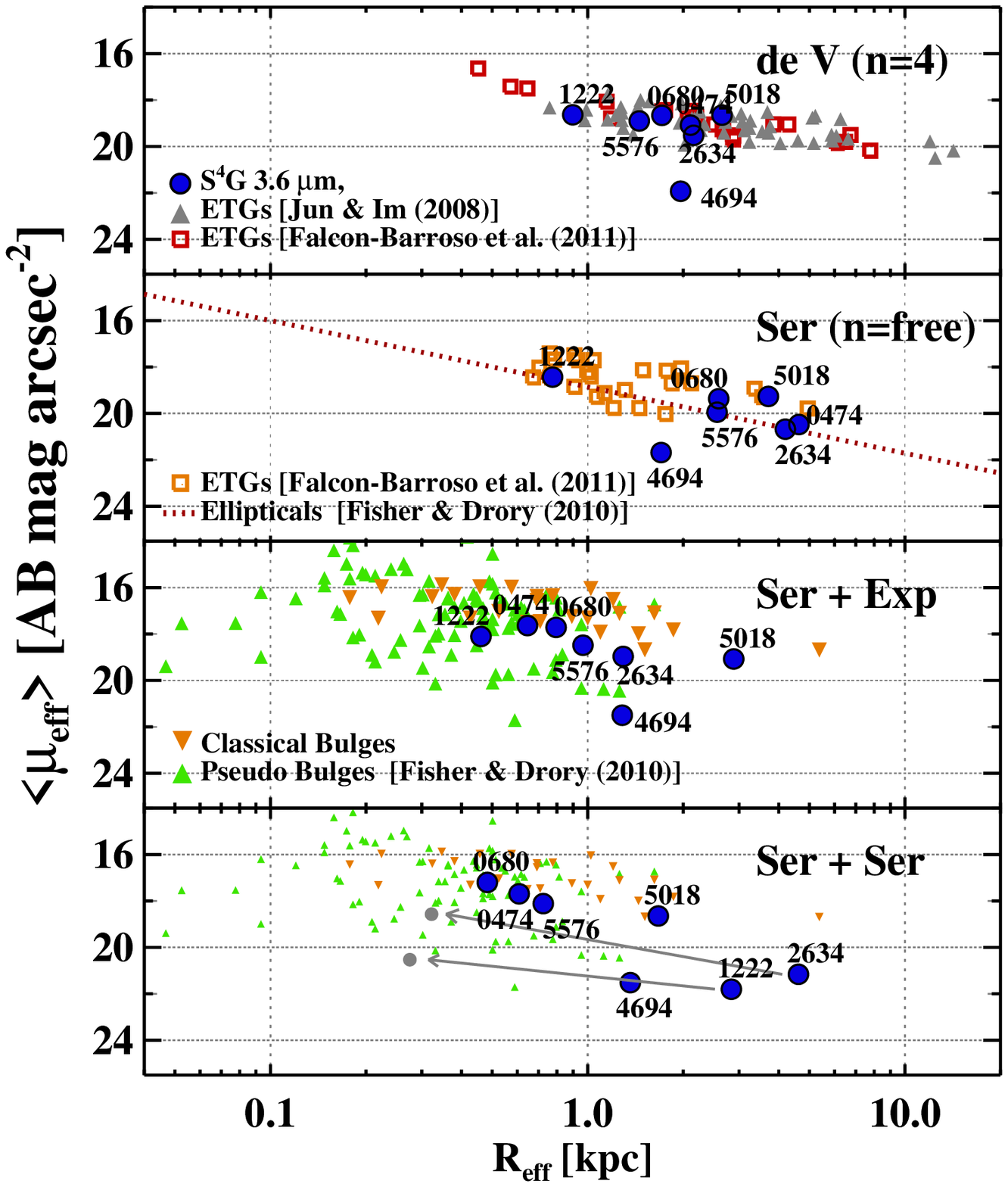}
\caption{ Kormendy relation of ETGs showing tidal features. The effective radii were converted to the circularized effective radii, $R_{\rm eff}$.  The uppermost panel shows the result of de Vaucouleur fit. Grey triangles indicate normal ETGs from \citet{jun08} and red squares indicate ETGs modeled with {\ser} index n=4 from \citet{falcon_barroso11}. The upper middle panel shows the result from n=free {\ser} fit. Red dotted line represents ellipticals of the Virgo cluster from \citet{kormendy09}. Orange squares indicate ETGs that were modeled with free {\ser} index from \citet{falcon_barroso11}.  
The lower middle panel shows the result of {\ser} (for the bulge component) plus exponential fit (for the disk component). Orange and green triangles represent the galaxies with classical and pseudo bulge, respectively from \citet{fisher10}. The lowermost panel shows the result of the fit with two {\ser} functions. Grey arrows from NGC 1222 and NGC 2634 connect to the points if we exchange the bulge and disk component (see text for detail).} \label{kormendy_rel}
\end{center}
\end{figure*}

\begin{figure*}
\begin{center} 
\includegraphics[width=14cm]{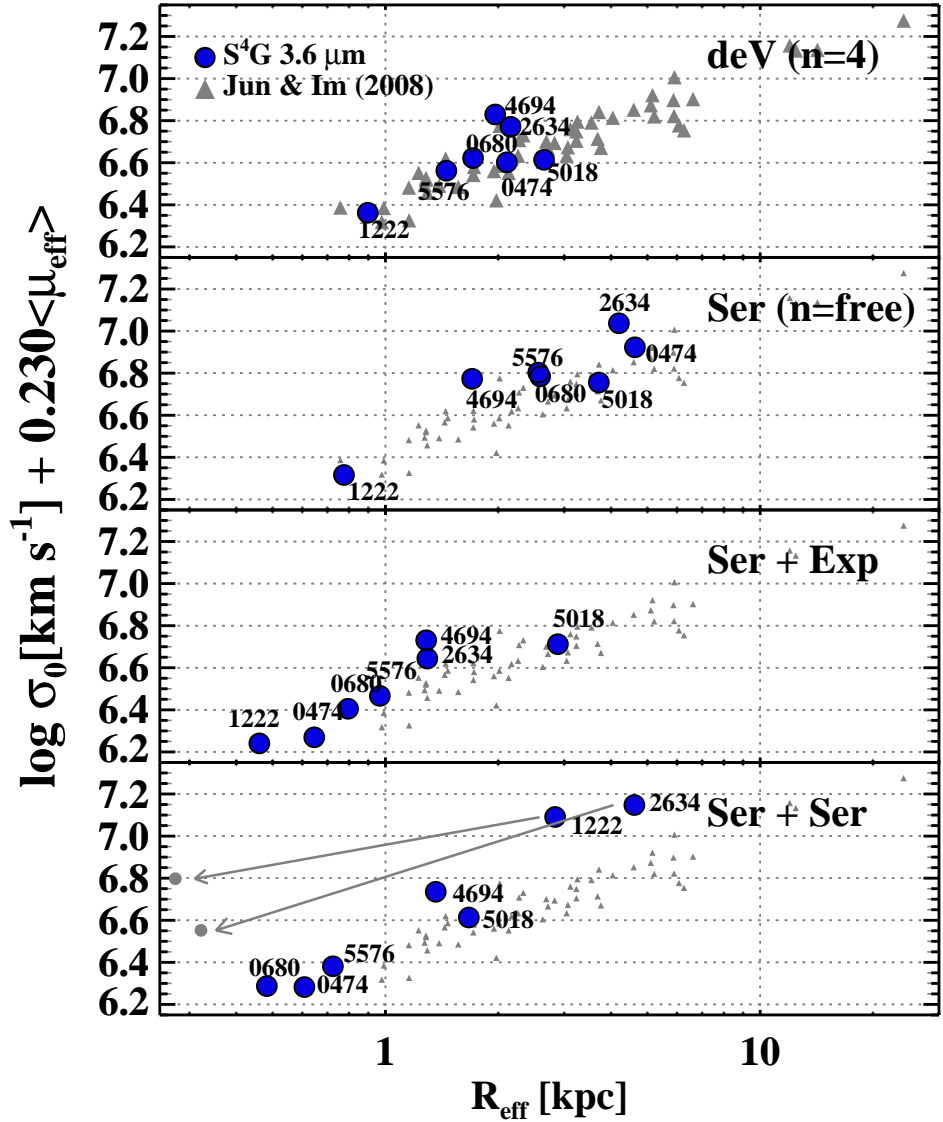} 
\caption{Fundamental Plane of ETGs with tidal features. The effective radii, which are presented here were converted to the circularized effective radii, $R_{\rm eff}$. Blue circles represent the result from this study and results from four different model fit were plotted in each panel. In the uppermost panel, blue circles represent the result of de Vaucouleur fit. ETGs of \citet{jun08} were plotted in grey triangles for comparison. 
The upper middle panel shows the result of {\ser} fit and the lower middle panel shows the result of {\ser} (for bulge) plus exponential (for disk) fit. $R_{\rm eff}$ and $\mu_{\rm eff}$ are those of the bulge component alone. The fit with {\ser}+{\ser} profile are plotted in the lowermost panel. Note that grey triangles of \citet{jun08} represent the result obtained with the $R^{1/4}$ profile, hence only the uppermost panel shows a direct comparison, but we plot them together in the other panels for comparison. Grey arrows from NGC 1222 and NGC 2634 connect to the points if we exchange the bulge and disk component (see text for detail).
\label{FP}}
\end{center}
\end{figure*}

Surface brightnesses and effective radii are determined by the choice of the adopted model profile, as evident in Figure \ref{compare_st}, \ref{kormendy_rel} and \ref{FP}. Therefore, we need to be careful when we compare our results with those from other studies, especially when we compare with galaxies that were modeled with different types of fit. We present the KR, a relation between the effective radius and the mean surface brightness at that radius for ETGs showing tidal features in Figure \ref{kormendy_rel}. 
To make comparisons with other studies, we also plot KRs from the literature, which were obtained from the same 3.6 {\micron}-band. 
Each panel shows the KR obtained with a specific model fit. 
We show the KR derived from the n=4 profile fit in the uppermost panel together with the results of ETGs from \citet{jun08} in grey triangles and \citet{falcon_barroso11} in red squares. Note that structural parameters ($r_{\rm eff}$ and $<\mu_{\rm eff}>$) of these two studies were obtained with one-dimensional {$R^{1/4}$} profile fittings, while our data were obtained with two-dimensional model fits. ETGs presented in \citet{falcon_barroso11} can be divided into two groups that have {\ser} index n=4 and n=free. 
We only plot ETGs that have n=4 in this panel.  In the second panel of Figure \ref{kormendy_rel}, we plot the result from a {\ser} (n=free) fit. To compare with normal elliptical galaxies, we draw a red dotted line adopted from Figure 8 of \citet{fisher10}, which represents elliptical galaxies of the Virgo cluster from (\citealt{kormendy09}).
Since these galaxies were originally studied in the optical band, \citet{fisher10} shifted the surface brightness measurement to the L'-band. We also plot ETGs from \citet{falcon_barroso11} that were fit with {\ser} index n=free in orange squares.
In the third panel, we show the result of a {\ser} (for the bulge) plus exponential fit (for the disk). In the {\ser}$+$exponential fit, the values of $r_{\rm eff}$ and $\mu_{\rm eff}$ are those of the bulge component. We also plot the classical bulge (in orange triangle) and pseudo bulge (in green triangle) of \citet{fisher10}. Their structural parameters were determined by fitting the surface brightness profiles with a one-dimensional {\ser} function plus an exponential outer disk. The last panel shows the result of the fit with a {\ser} plus {\ser} function. The classical and pseudo bulges are the same as above, and they were plotted just for comparison. However, because a different fitting model is used, we can not apply direct comparison in the lower most panel. NGC 4694 is a galaxy with a nuclear HII region (\citealt{veron-cetty06}) and the velocity dispersion of this galaxy is 61 \kms. Therefore it falls below the main KR of ETGs, similar to dwarf galaxies showing nuclear star formation. Except for NGC 4694, ETGs with tidal features do not stand out in the KR from the non-interacting normal ETGs of \citet{jun08} and \citet{falcon_barroso11}.  
  
We present the FP, which is an empirical scaling relation between the effective radius, the mean surface brightness, and the central velocity dispersion of tidally disturbed ETGs in Figure \ref{FP}, where the axes are projected in the direction of the smallest scatter at $3.6$ {\micron} as provided by \citet{jun08}. As in Figure \ref{kormendy_rel}, we show FPs derived with four different surface brightness profile models. In this figure, we compare the FPs of our sample galaxies only with those of \citet{jun08}. The triangles of Jun \& Im (2008) represent the results obtained with the $R^{1/4}$ profile, hence only the uppermost panel shows a direct comparison, but we plot them together in the other panels for comparison. Velocity dispersions were adopted from Hyperleda. 
As discussed in the previous section, the {\ser}$+${\ser} model fit of NGC 1222 and NGC 2634 may not be ideal because the effective radii of the bulge component are larger than those of the disk component. So, we draw grey arrows in the lowermost panel of Figure \ref{kormendy_rel} and \ref{FP} representing an exchange of the values of r$_{\rm eff}$ and $\mu_{\rm eff}$ of the bulge and disk components.
We find that tidally disturbed ETGs lie on the same plane as do normal ETGs (\citealt{jun08}) and Virgo ellipticals (\citealt{kormendy09}). Because galaxies exhibiting tidal features do not occupy a distinct region in the KR or FP, this implies that the structural properties of the spheroidal components of tidally disturbed galaxies are not significantly different from those of the other ellipticals or the spheroids of galaxies with no signs of tidal interaction.

%---------------------------------------------

Our result is consistent with the findings of numerical simulations of mergers that show that the FP is preserved during gas-poor, dissipationless major merging events (\citealt{capelato95}; \citealt{gonzalez03}; \citealt{nipoti03}; \citealt{boylan05}; \citealt{boylan06}). \citet{rothberg06} also find observationally that most of the single-nuclei merger remnants, including shell galaxies, appear to lie on the $K$-band FP or within the scatter of the ellipticals taken from \citet{pahre99}.

On the other hand, simulations show that projections of the FP, such as the KR and the Faber-Jackson relation (\citealt{faber76}, FJR), are not preserved via dry mergings (\citealt{nipoti03}; \citealt{boylan06}). \citet{boylan06} showed that the location of merger remnants in the FJR and KR is sensitive to the amount of the energy transferred from the bulge to the dark matter haloes during the merger. 
They claimed that mergers on orbit with large angular momentum would experience more dynamical friction. This results in galaxies with larger velocity dispersion (smaller $\beta$ in $R_{\rm eff}  \propto  L \sigma^{\beta}_{\rm eff}$) and smaller r$_{\rm eff}$ (smaller $\alpha$  in $R_{\rm eff}  \propto   M^{\alpha}_{*}$). Therefore, such galaxies form a less steep KR and FJR. 
While the end product of mergers on orbits with less angular momentum will form steeper KR and FJR, i.e., have higher $\alpha$ and $\beta$.
But in Figure 5 (KR) of \citet{rothberg06}, shell galaxies and other merger remnants, except for luminous and ultra-luminous infrared galaxy mergers, also lie within the scatter (\citealt{pahre99}), consistent with our result in the KR.

Our results do not mean that the properties of the galaxies showing tidal debris have not changed as a result of the merging events. It is more likely that tidal features induced in minor mergers or interactions, do not affect bulge properties. Alternatively if those tidal features were caused by major mergers, then enough time has passed for their bulges to relax and become similar to those of ETGs that do not show tidal disturbances. By checking H$\beta_{0}$ line indices, \citet{falcon_barroso11} show that outliers from the FP tend to have young stellar populations in common. Having young stellar populations might be a more efficient way to push galaxies off the FP than dry merging as also suggested by the E+A findings (\citealt{yang08}). We infer that the dynamical scattering of the FP due to a merging event last only for a relatively short time scale. This is also in agreement with the finding that ETGs show a tight relation in the FP. Thus through dry merging events between galaxies which do not have enough gas to trigger star formation, galaxies appear to slide along the FP in the direction of larger effective radii, rather than lie off the FP.

\section{Summary and Conclusions}
In a sample of ETGs drawn from \s4g that have $T \le 0 $, we perform unsharp masking, two-dimensional fitting, and derive structure maps to search for tidally disturbed ETGs. 
We identify shells, ripples, tidal tails or broad fans of stars in $17\pm3\% $ (11/65) of ETGs with the imaging data that have average depth of $<\mu_{3.6,AB}> (1\sigma) \sim$ 27 mag arcsec$^{-2}$ at $3.6$ {\micron}. 

We estimate that the tidal disturbances contain $3$ -- $10\%$ of the total 
3.6 {\micron} galaxy luminosity, which can be directly translated to the stellar mass contained in the debris.

To investigate the structural properties of tidally disturbed galaxies, we modeled the galaxies with four different functions, de Vaucouleurs, {\ser}, {\ser}+exponential and {\ser}+{\ser} profile. We find that the precise measurement of structural parameters depends on the adopted decomposition model. 
We also find that the total magnitudes of galaxies vary among different model fits.

We explored the position of these galaxies in the Kormendy relation and Fundamental Plane and find that ETGs with tidal debris occupy the same area as those galaxies without tidal debris or bulges of early type systems. This implies that the bulge properties of tidally disturbed galaxies are not significantly different from those of normal ETGs. We infer that dry merging cause galaxies to slide along the FP and minor merging or weak interactions that cause tidal disturbances in our sample either affect only outer part of the galaxy, or enough time has elapsed that the bulges of the tidally disturbed galaxies are well relaxed. Thus, having young stellar populations might be a more efficient way to make galaxies deviate from the FP than going through dry merging events.

The \s4g will survey over 2,300 nearby galaxies by the end of 2012, and this study will be extended to include a complete set of 180 ETGs.

\acknowledgments
{\it Facilities:} \facility{ The {\em{Spitzer}} Space Telescope}
T.K. is grateful to the entire \s4g team for their efforts in this project. We thank Hyunsung David Jun for kindly sharing the data of Jun \& Im (2008) paper. We also thank Jesus Falcon-Barroso for providing us with updated data of \citet{falcon_barroso11}.  We are grateful to the anonymous referee for helpful comments and suggestions that have improved this paper.
 
 T.K., K.S., J.-C.M.-M., and T.M. acknowledge support from the National Radio Astronomy Observatory, which is a facility of the National Science Foundation operated under cooperative agreement by Associated Universities, Inc. M.G.L. was supported in part by Mid-career Research Program through the NRF grant funded by the MEST (no.2010-0013875). This work was co-funded under the Marie Curie Actions of the European Commission (FP7-COFUND). E.A. and A.B. thank the thank the Centre National d'Etudes for support.
  
This research is based on observations and archival data made with the {\em{Spitzer}} Space Telescope, and made use of  the NASA/IPAC Extragalactic Database (NED) which are operated by the Jet Propulsion Laboratory, California Institute of Technology under a contract with National Aeronautics and Space Administration (NASA). We acknowledge the usage of the HyperLeda database (http://leda.univ-lyon1.fr).
\bibliography{tkim_etg}  

%-------------------------------------------------------

\end{document}